%% file: manuscript.tex
\documentclass[10pt,leqno]{amsart}
\usepackage{graphicx}
\usepackage{upgreek}
\usepackage{multicol,multirow}
\usepackage{amsmath,amssymb,amsfonts}
\usepackage{mathrsfs}
\usepackage{amsthm}
\usepackage[figuresright]{rotating}
\usepackage{appendix}
\usepackage[numbers]{natbib}
\usepackage{ifpdf}
\usepackage[T1]{fontenc}
\usepackage{newtxtext}
\usepackage{newtxmath}
\usepackage{textcomp}
\usepackage{xcolor}
\usepackage[singlelinecheck=false,labelfont=it,textfont=normalfont,singlelinecheck=off,justification=raggedright]{subcaption}
\usepackage{amsmath,amsfonts, empheq}
\usepackage{systeme}
\usepackage{amssymb}
\usepackage{wasysym}
\usepackage{lscape}
\usepackage{wrapfig}
\usepackage{siunitx}
\usepackage{comment}
\usepackage{booktabs, array}
\usepackage{tabularx}
\usepackage{tikz}
\usetikzlibrary{automata, arrows.meta, positioning}
\usetikzlibrary{arrows,cd}
\usepackage{amsmath}
\usepackage{breqn}
\usepackage{siunitx}

\usepackage[colorlinks,allcolors=blue]{hyperref}
\definecolor{jourcolor}{cmyk}{1,0.57,0.01,0.38}
\hypersetup{
    colorlinks,%
    citecolor=jourcolor,%
    filecolor=jourcolor,%
    linkcolor=jourcolor,%
    urlcolor=jourcolor
}

\theoremstyle{definition}

\newcommand{\rd}{\mathrm{d}}

\begin{document}

\title[Mixing and ventilation in a living laboratory due to fast and slow response modes]{Mixing and ventilation in a living laboratory due to fast and slow response modes}

\author[Costanza Rodda, John Craske, and Graham O. Hughes]{Costanza Rodda$^{\ast}$,John Craske and Graham O. Hughes}

\address[1]{Department of Civil and Environmental Engineering, Imperial College London, SW7 2AZ, UK}

\maketitle

\begin{abstract}
We present and analyse observational data from a highly instrumented classroom computer laboratory and develop a multizone model to describe its mechanical ventilation and mixing regime. The laboratory houses 70 workstations that are used heterogeneously in time and space, in a manner similar to a generic office environment. Our model predicts CO$_2$ concentration in the laboratory, accounting for air exchange between the occupied classroom and its ceiling plenum and by parametrising irreversible mixing in each zone. 
Applying the model to our measurements helps identify critical components in the ventilation network, as highlighted by a strong separation of the time scales characterising the flow response. On the one hand, this time scale separation leads to a simplified model describing the CO$_2$ transport. On the other hand, it suggests that the forced exchange of volume between the room and the plenum is `overdriven' in that reduced energy operation could be achieved without compromising air quality. More generally, our modelling approach offers a systematic method to enhance energy efficient ventilation of multi-zone systems.
\end{abstract}

\section{Introduction}

\subsection{Motivation}
The COVID-19 pandemic and the unprecedented energy crisis many countries have faced recently have highlighted the need for new measures that maximise energy efficiency and simultaneously ensure a safe and comfortable environment for building occupants. Ventilation is recognised as an essential ongoing measure to reduce the risk of the indoor spreading of airborne infectious diseases, such as the SARS-CoV-2 virus \cite{bhagat2020effects, bazant2021guideline, burridge2022predictive}. During the COVID-19 pandemic, many countries opted to increase the running time of Heating, Ventilation, and Air Conditioning (HVAC) systems, following recommendations from several studies \cite{dai2020association, sun2020efficacy, guo2021review}. However, HVAC systems are among the main contributors to a building's life cycle carbon emissions and impact heavily upon its running costs because they are associated with approximately 75\% of the total energy consumption \cite{laustsen2008energy}. 

Better design and control of HVAC systems require the development of relatively simple models that can be used efficiently at the scale of an entire building. This is challenging from a fluid mechanics perspective because the transport and mixing of air in buildings is spatially heterogeneous and occurs on a wide range of temporal and spatial scales.

\subsection{Existing models}
Several different approaches have been developed in the past decades to optimise HVAC control and improve the energy efficiency of systems \cite{afram2014theory}. One such approach is demand-controlled ventilation (DCV), which uses feedback control methods to optimise the ventilation rate based on building occupancy variations. DCV has led to a substantial reduction in energy consumption (up to 45\% \cite{pang2020nationwide}) because a significant amount of energy would be otherwise wasted during unoccupied and low-occupation hours \cite{masoso2010dark}.
Occupancy measurement relies on either direct counting or an indirect proxy - typically CO$_2$ concentration.
Advances in direct counting have been enabled by technologies such as wireless technologies (WiFi), Bluetooth, camera-based systems, and environmental sensors \cite{zhao2022state}. Nevertheless, significant errors remain with most of these techniques.

CO$_2$ concentration-based occupancy measurement relies on the strong correlation between CO$_2$  levels and the number of occupants. This correlation can be calculated for a single-zone space under the following assumptions: the air in the zone is well-mixed at all times and the occupants are the only source of CO$_2$. If these assumptions hold true, CO$_2$ concentration can be related to occupancy using a simplified mass-balance equation

\begin{equation}
V \frac{dC}{dt} = -QC+NF,
\label{eq:mass-balance}
\end{equation}
where $V$ is the volume and $C$ is the CO$_2$ concentration above the outdoor concentration (under the assumption that the outdoor concentration does not change with time), $Q$ is the ventilation rate, $N$ is the number of people and $F$ is the CO$_2$ generation rate per person \cite{lu2022nexus}. 

In spaces containing multiple zones, transient airflows and complex interactions between zones can significantly affect CO$_2$ distributions and hence occupancy estimates.

\subsection{Overview}

In this work, we propose a technique that combines modelling and data analysis to address the challenge posed by the coupled problem of energy usage reduction and air quality control. 

First, we derive a mass-balance equation for an arbitrary zone (sub-volume) within a building. Irreversible mixing in the zone is parameterized and our model makes explicit use of the physics that couples CO$_2$ and occupancy.

Then, we apply a dynamical systems approach to analyse the complex behaviours of airflows in multi-zone spaces. By utilising the spectral properties of the system, we identify response timescales that may be considered  ``fast'' and ``slow''. This timescale separation enables us to project the system onto a lower-dimensional space, allowing for more effective analysis and optimisation. The use of timescale separation to simplify the equations describing a system and understand its properties has been used at length in different contexts, such as weather modelling \cite{pena2004separating}, geophysical flows \cite{vanneste2013balance}, combustion chemistry \cite{al2009one} and homogenisation for stochastic modelling more generally, where it has also been referred to as `adiabatic elimination' \cite{PavGboo2008a,HakHboo1983a}. However, its application in the analysis of building ventilation has not been exploited.

Finally, we examine the model performance against the real data, providing insights into the impact of various parameters on CO$_2$ concentrations and potential energy savings.

Parker \& Bowman \cite{parker2011state} have applied a state-space formulation to study the analytical solutions and characterise the dynamical behaviour of multi-zone buildings in the context of contaminant transport. We adopt a similar approach, investigating the importance of key parameters in relation to the characteristics of our system. In this study we demonstrates how our method can be applied to ventilation systems. Geometrically, we represent the ventilation response as trajectories in phase space, which proves to be an insightful way of characterising the system's behaviour. 

To illustrate the application of this approach in understanding and modelling air exchanges among connected spaces and parameterizing the mixing of air within sub-spaces, we analyse observational data from a `living' laboratory -- a highly instrumented classroom at Imperial College London. The laboratory is representative of a wide range of mixed ventilation spaces with characteristics encompassing many building types, including offices (with not less than 6 m$^2$/person), which are ubiquitous worldwide. The data measured in the laboratory provide a way to understand how the system evolves in response to different configurations and forcings. The model of our living laboratory is represented by a space composed of a plenum containing fan coil units above the suspended ceiling of the occupied room -- a typical configuration for mechanically ventilated rooms. In this space, we consider carbon dioxide (CO$_2$) as a proxy for ventilation and respiratory contaminants. CO$_2$ can be modelled relatively easily since its transport is limited to advection (or ventilation), while other quantities like heat are more complicated because transport occurs by convection, conduction, and radiation and the sources are comparatively difficult to quantify.

The laboratory, measurements, and dataset are introduced in \S\ref{sec:method}, and the analytical model is described in \S\ref{sec:theory}. Section \ref{sec:model-data} focuses on comparing the model with the observed data, and our discussions and conclusions are given in \S\ref{sec:discussion} and \S\ref{sec:conclusions}.

\section{A `living' laboratory and dataset}
\label{sec:method}
This study is based on data measured in a classroom equipped with a cluster of computer workstations--a `living' laboratory--located at the South Kensington Campus of Imperial College London. Figure \ref{fig_room} shows a 3D sketch of the laboratory highlighting its main features. The laboratory consists of two spaces: a main room hosting the occupants and workstations (20.1 m in length, 8.0 m in width, and 3.2 m in height for a total volume $V_0 \approx 515$ m$^3$) and a plenum situated above the room hosting the air conditioning system (20.1 m in length, 8.0 m in width, and 0.5 m in height for a total volume $V_1 \approx 80$ m$^3$). The laboratory is adjacent to a lecture room on the west side and an open-plan student space on the north and east sides. The south-facing facade has 12 double-glazed windows, each with a surface area of \SI{2.6}{\meter\squared} that are locked in a closed position and cannot be opened by the occupants. The laboratory is accessible through two doors located on the north wall. These doors are self-closing (and alarmed) and, therefore, the period for which the doors are open (while people pass into and out of the room) is assumed not to affect the room interior. The laboratory is subject to an extensive range of spatially and temporally heterogeneous thermal forcing; it hosts 70 workstations, a similar number of potential occupants (floor space per person therefore ranging from approximately 2 to 160 m$^2$), and is mechanically heated and cooled (see \S\ref{sec:ventilation}). The laboratory has been instrumented throughout with numerous CO$_2$, humidity and temperature sensors (see \S\ref{sec:measurements}).%

\begin{figure}[t]
\centering
\scalebox{0.8}{\input{diag_v2}}
\caption{Schematic diagram of the computer laboratory. The coloured arrows and lines indicate the airflow directions occurring within the ducts in the plenum. The blue lines/arrows indicate the supply of air from the air handling units that are injected into the room by the four FCUs. The red lines/arrows indicate air that has been extracted from the room through the five ceiling grilles. The CO$_{2}$ sensors are marked $S_{ph}$, where `$p$' corresponds to the (horizontal) position and the index `$h$' corresponds to height, for which the integers $1$, $2$, $3$ and $4$ correspond to 3m, 2m, 1m and 0.5m, respectively, above floor level.}
\label{fig_room}
\end{figure} 
%

\subsection{Heating ventilation and air conditioning (HVAC)}
\label{sec:ventilation}
The laboratory is mechanically ventilated via a typical Variable Air Volume (VAV) system controlled by a Building Management System (BMS). Temperature-controlled fresh air is pumped into a plenum above the room (see the blue pipe in figure \ref{fig_room}). Four Fan Coil Units (FCUs) (Quartz Sapphire model SPR9) draw air from the plenum and heat/cool the air using a hot/cold water system according to the temperature setpoint for the room. The conditioned air is ducted from the FCUs to supply the room via linear slot diffusers placed at the north and south sides of the room (blue pipes and arrows in figure \ref{fig_room}). The rate of supply to the room is usually different to the rate of extraction, which is constant and occurs through five square grilles positioned along the longitudinal axis of the ceiling (red pipes in figure \ref{fig_room}). 

%
To account for the possible mismatch between the rate at which air is supplied by the FCUs and extracted according to the VAV system,
the plenum and the room exchange air through two additional grilles on the ceiling (not shown). In addition, sections of the linear slot diffusers are not connected to the FCUs and allow air exchange
between the room and the plenum. These air exchanges can occur in
either direction (from the room towards the plenum and vice versa)
depending on the AHU regime. 

\subsection{Occupancy}
\label{sec:occupancy}
Occupancy estimates are inferred from Imperial's deployment of the HubStar (formerly LoneRooftop) Building Insights Dashboard, which records the number of WiFi connections (hourly average and maximum values are available).
WiFi estimated occupancy has several advantages, such as protecting the privacy of occupants and achieving a high accuracy ($96\%$ according to the study by Simma et al.\ \cite{simma2019real}). These advantages and the wide availability of WiFi signals in buildings have made it a convenient means of occupancy estimation in many studies in the past ten years \citep{zhao2022state}. Nevertheless, estimates have uncertainties associated with the varying number of WiFi devices each person carries. In our case, the time resolution and the anonymisation of the recording of WiFi connections leads to the estimated uncertainty of the order of 30\% (see Appendix \ref{sec:uncertainty} for more details).

\subsection{Measurements}
\label{sec:measurements}
The laboratory is monitored continuously with 24 temperature
sensors (Trend thermistors model T/TFR 4) and 8 combined
CO$_2$/temperature/humidity sensors (TREND space sensors model
RS-WMB-THC) connected to the BMS (via 4 Trend IQ4 controllers). The temperature
sensors are installed on 4 vertical risers (i.e. 6 per riser) at
different positions in the room. The combined sensors are situated
on vertical walls at different horizontal positions in the room (see \ref{sec:measurements} for more details). The CO$_2$ and
temperature measurements were further supplemented by sensors in the supply and extract ducting of the BMS system.

For the present study, we focus on the data provided by the 8 combined CO$_2$/temperature/humidity sensors. The dataset spans from January 2022 to September 2022, with measurements sampled and recorded every 5 minutes.
We assume that the occupants are the only source of CO$_2$ in the room.

We supplement the data acquired in the room with a range of standard measurements recorded by the BMS at 5-minute intervals. These measurements include the supply and extraction volume flow rates, $Q_{\text{in}}$ and $Q_{\text{out}}$, respectively and the supply and return temperatures from the FCU. The flow rates $Q_{\text{in}}$ and $Q_{\text{out}}$ are measured by sensors placed in the air ducts. The data are presented in \S\ref{sec:model-data}.

\section{Analytical model}
\label{sec:theory}
\begin{figure}[t]
  \begin{subfigure}[t!]{0.65\textwidth}
    \centering
    \caption{}
    \input{plenum}
  \end{subfigure}
  \begin{subfigure}[t!]{0.35\textwidth}
    \centering
    \caption{}
\begin{tikzcd}[column sep=huge, row sep=large]
 0\arrow[dr, "(1-\gamma_{1})q"]\arrow[d, "\gamma_{1}q"] & \\
 C_{1}\arrow[r, "\gamma_{1}q+q'"] & C_{1}'\arrow[d, "\gamma_{0}Q"]\arrow[dl, "(1-\gamma_{0})Q"] \\
 C_{0}'\arrow[u, "q'"']\arrow[uu, bend left=50, "q"] & C_{0}\arrow[l, "\gamma_{0}Q"] 
\end{tikzcd}
\end{subfigure}
\caption{$(a)$ Schematic diagram (not to scale) of a vertical section through the ceiling plenum and computer laboratory below, with excess concentrations $C_1$ and $C_0$, respectively; the corresponding control volumes are highlighted with a red dashed line. $C'_1$ and $C'_0$ indicate the excess CO$_2$ concentration in the FCU and ceiling zone (highlighted with a darker shade of grey). As illustrated by the white arrows, conditioned air is supplied to the plenum with a volume flux $q$. A fraction $\gamma_{1}\in[0,1]$ of that supply air mixes with the air within the plenum; the rest is drawn directly into the fan coil unit (FCU), whose fan drives a total volume flux $Q=q+q'$. After being fed into the computer laboratory, a fraction $\gamma_{0}\in[0,1]$ of air from the FCU either mixes with the air in the room; the rest comprises a `short circuit' and remains in the ceiling zone. Due to the fact that $Q\neq q$, the FCU drives a secondary volume flux $q'=Q-q$ between the computer laboratory and plenum. $(b)$ Graph corresponding to the governing equations described in \eqref{sec:theory}. Each node corresponds to a control volume (labelled with CO$_2$ concentration) and each branch corresponds to a flux of $CO_{2}$ between control volumes (labelled with the corresponding volume flux).}
\label{fig_sketchSide}
\end{figure}

\subsection{CO$_2$ budget}
\label{sec:model}
Let $C_{0}$ and $C_{1}$ be the bulk excess CO$_2$ concentration (relative to outdoors) in the room
and plenum, respectively, and let $C_{0}'$ and $C_{1}'$ be the
quasi-steady state excess concentration associated with the ceiling zone
and FCU, respectively (see figure \ref{fig_sketchSide} for a schematic representation of the spaces in a vertical cross-section of the room). We use $q$ and $q'$ to denote the rates at which fresh air is supplied to the plenum and at which air is drawn into the plenum from the ceiling
zone, respectively, and define $Q:=q+q'$ (see figure \ref{fig_sketchSide}). The variable $Q$, therefore, corresponds to the volume flux driven by the FCUs.

First, we establish an equation for the CO$_2$ concentration in the FCUs, the combined volume of which is labelled $V_{1}'$. Unless stated otherwise, all CO$_2$ concentrations given refer to excess values.

Since the supply air duct is not connected directly to the FCUs, let $\gamma_{1}\in [0,1]$ represent the fraction of the supply rate $q$ that first mixes with the air in the plenum before entering the FCU:

\begin{equation}
  V_{1}'\frac{\rd C_{1}'}{\rd t}=\underbrace{(\gamma_{1}q+q')C_{1}}_{\mathrm{from\ plenum}} -\underbrace{QC_{1}'}_{\mathrm{to\ room}},
\label{eq:C1p}
\end{equation}
for which we have assumed that $\gamma_{1}q+q'\geq 0$, i.e.\ the FCU always extracts air from the plenum.

For the control volume just below the room's ceiling, we adopt a similar approach to the FCU, using $\gamma_{0}\in [0,1]$ to denote the fraction of FCU outlet air that first mixes with the rest of the room (before entering the ceiling zone):
\begin{equation}
  V_{0}'\frac{\rd C_{0}'}{\rd t}= \underbrace{\gamma_{0}QC_{0}}_{\mathrm{from\ room}}+\underbrace{(1-\gamma_{0})QC_{1}'}_{\mathrm{from\ FCUs}}-\underbrace{q'\phi}_{\mathrm{to\ plenum}} -\underbrace{qC_{0}'}_{\mathrm{to\ extract}},
\label{eq:C0p}
\end{equation}

\noindent where

\begin{equation}
\phi=\begin{cases}
C_{0}' &\quad q'\geq 0, \\
C_{1} &\quad q'<0,
\end{cases}
\end{equation}
\noindent allows the flux of CO$_{2}$ to be represented as  either from the ceiling zone to the plenum
($q'\geq 0$) or from the plenum to the ceiling zone ($q'< 0$). 

Let us now consider the CO$_2$ concentration in the room and plenum. We assume that the rate of CO$_2$ generation per person is $F = 0.012$ g/s/person, consistent with the range $F = 0.009-0.012$ g/s/person  for occupants aged 21 to 30 years undertaking typical office work \cite{persily2017carbon}. Indeed, our data (to be discussed in Figure \ref{fig_timedata}) will indicate that $F = 0.012 \pm 0.001$ g/s/person provides the best fit for the model, as evaluated by the highest $R^2$ value. Furthermore, we assume that the infiltration and exfiltration of air are negligible (the validity of such an assumption is discussed in \S\ref{sec:inf_exf}). The system of linear ODEs describing the time-variation of concentration in the room and the plenum, respectively, is therefore
\begin{align}
V_{0}\frac{\rd C_0}{\rd t} & = \overbrace{\gamma_{0}QC_{1}'}^{\mathrm{from\ FCU}}-\overbrace{\gamma_{0}Q C_{0}}^{\mathrm{to\ ceiling\ zone}}+NF, \label{eq:room}\\
V_{1}\frac{\rd C_1}{\rd t} & = \underbrace{q'\phi}_{\mathrm{from\ ceiling\ zone}} - \underbrace{(\gamma_{1}q+q')C_{1}}_{\mathrm{to\ FCU}}, \label{eq:plenum}
\end{align}
where $N$ is the number of occupants. The system above is similar to the two-zone transient model proposed by \cite{lawrence2006evaluation}, with the difference that we have relaxed the condition of complete mixing in the plenum ($V_1$) and occupied zone ($V_0' + V_0$) by adding the $\gamma$ parameters.
The governing equations can be graphically represented in the network diagram in figure \ref{fig_sketchSide}(b), where the nodes represent control volumes labelled with concentrations. The volume fluxes between control volumes are represented by the network's branches.

Let us now re-write the governing equations \eqref{eq:C1p}, \eqref{eq:C0p}, \eqref{eq:room}, and \eqref{eq:plenum} in non-dimensional form by introducing the following variables:
\begin{equation}
  \tau := \frac{q}{V_1}t, \quad \varepsilon := \frac{V_1}{V_0}, \quad \zeta:= \frac{q'}{q}, \quad f_{0} := \frac{NFV_1}{q V_0}.
  \label{eq:adim}
\end{equation}
Physically, $\tau$ is a dimensionless time based on the filling time scale associated with the plenum (in terms of its volume $V_{1}$ and the AHU supply volume flux $q$), $\varepsilon$ represents the ratio of the plenum and room volumes, $\zeta$ is the ratio of the rates of room-plenum re-circulation and ambient fresh air introduction, and $f_0$ is the dimensionless forcing.

We further assume that the FCUs and ceiling buffer zone volume are much smaller than the plenum volume, i.e.\ $V_1'/V_1 \ll 1$ and $V_0'/V_1 \ll 1$.
In physical terms, storage of CO$_2$ in the FCU and ceiling buffer zone, represented by the time derivative in \eqref{eq:C1p}-\eqref{eq:C0p}, is therefore assumed to be unimportant, such that \eqref{eq:C1p} and \eqref{eq:C0p} give explicit algebraic expressions relating $C_0'$ and $C_1'$ to $\gamma_0$, $\gamma_1$, $q$ and $q'$. 

Under these assumptions, the resulting non-dimensionalised equations are

\begin{align}
\frac{\rd C_0}{\rd \tau} & = - \varepsilon\gamma_0(\zeta+1) C_0 + \varepsilon\gamma_0(\zeta+ \gamma_1)C_1 + {f_0} \label{eq:room_a}, \\ 
\frac{\rd C_1}{\rd \tau} & = \zeta \phi - (\zeta + \gamma_1)C_1. 
\label{eq:plenum_a}
\end{align}
We can express the system (\ref{eq:room_a}) and (\ref{eq:plenum_a}) in the form
\begin{equation}
 \frac{\rd\boldsymbol{C}}{\rd\tau} = \boldsymbol{A}\boldsymbol{C}+\boldsymbol{f},
\label{eq:x}
\end{equation}
where $\boldsymbol{A}$ is a coefficient matrix, $\boldsymbol{C} = (C_0, C_1)^{\top}$ is the state vector and $\boldsymbol{f}=(f_0, 0)^{\top}$ is the forcing vector. 

Assuming that $\boldsymbol{A}$ is diagonalisable (see discussion in \S\ref{sec:modelMoredim} for the general case in which this assumption is not valid), we can derive the solution for \eqref{eq:x}
\begin{equation}
\boldsymbol{C} = \boldsymbol{R}\exp(\boldsymbol{\Lambda} \tau)\boldsymbol{R}^{-1}(\boldsymbol{C}(0) +\boldsymbol{A}^{-1}\boldsymbol{f})-\boldsymbol{A}^{-1}\boldsymbol{f}.
\label{eq:Csol}
\end{equation}
where $\boldsymbol{C}(0)$ are the initial conditions, $\boldsymbol{\Lambda}$ is the eigenvalue matrix and $\boldsymbol{R}$ is the corresponding (right) eigenvector matrix. The derivation of \eqref{eq:Csol} and explicit expressions for the eigenvalues and eigenvectors are provided in Appendix \ref{sec:solution}.
%
\subsection{Phase space diagrams}
\label{sec:PSD}
\begin{figure}[t]
\centering
\includegraphics[width=1\linewidth]{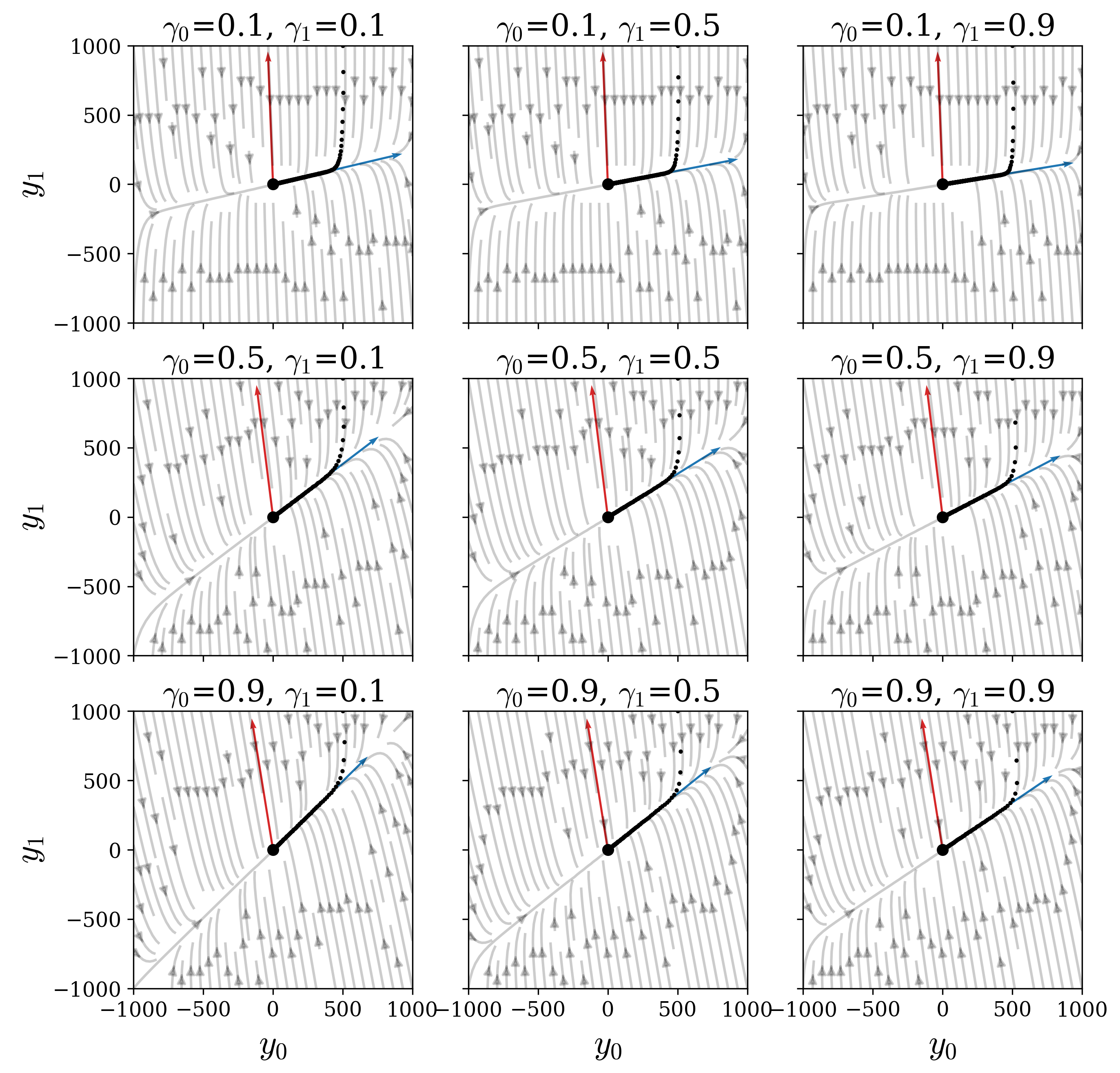}
\caption{Phase space diagram corresponding to the variable $\boldsymbol{y}$ (equation \eqref{eq:y}, which accounts for the stationary equilibrium $\boldsymbol{A}^{-1}\boldsymbol{f}$) for fixed $\varepsilon = 0.17$, $\zeta=1.8$, and several combinations of $\gamma_0$ and $\gamma_1$ values. The grey arrows show the flow evolution trajectories for different initial conditions, and the black dots show one particular solution of the dynamical system at uniform time intervals. The red and blue arrows show the two eigenvectors.}
\label{fig_phasetheory}
\end{figure} 
Each possible state of the system is represented by a point in phase space, and a sequence of consecutive states forms a trajectory. 
Therefore, phase space diagrams are useful for representing systems geometrically and investigating their equilibria, stability and evolution.

If the forcing $\boldsymbol{f}$ is regarded as constant then \eqref{eq:x} can be expressed as the autonomous system:

\begin{equation}
 \frac{\rd\boldsymbol{y}}{\rd\tau} = \boldsymbol{A}\boldsymbol{y},
\label{eq:odey}
\end{equation}

where 

\begin{equation}
\boldsymbol{y} := \boldsymbol{C}+\boldsymbol{A}^{-1}\boldsymbol{f}.
\label{eq:y}
\end{equation}

Figure \ref{fig_phasetheory} illustrates the phase space associated with \eqref{eq:odey}. We choose $\varepsilon = 0.17$ and $\zeta=1.8$ in \eqref{eq:room_a}-\eqref{eq:plenum_a} because they represent the measurements in the living laboratory and are characteristic of a ventilated office space. The values of $\gamma_0$ and $\gamma_1$ are varied to understand how the (parameterized) mixing affects the rate of convergence towards one manifold of the system.
The grey arrows in figure \ref{fig_phasetheory} mark flow evolution trajectories for different initial conditions. The red and blue arrows mark the directions of the two eigenvectors and correspond to the fast (large negative eigenvalue) and slow (small negative eigenvalue) dynamics of the system, respectively.

The black dots show one particular solution of the dynamical system at uniform time intervals. They show that the solutions converge rapidly towards the slow eigenmode, which dominates the adjustment to the equilibrium state $\boldsymbol{y}=\boldsymbol{0}$. 

In terms of the parameters, we can easily see that $\gamma_1$ has far less influence on the dynamics than $\gamma_0$. This behaviour comes from the condition $\varepsilon\ll 1$, which corresponds to the volume of the plenum being much smaller than the volume of the room below ($V_1 \ll V_0$). When $\gamma_0 \to 1$ and $\gamma_1 \to 0$, the CO$_2$ concentrations in the room and the plenum become comparable, as can be seen by the blue vector approaching a slope of one in the bottom right panel in figure \ref{fig_phasetheory}. When $\gamma_0 = 1$ and $\gamma_1 = 0$, the slow manifold maps onto $y_0=y_1$, while the fast manifold corresponds to the difference between the CO$_2$ concentration in the room and plenum. 
In all other cases shown in figure \ref{fig_phasetheory}, the concentration is higher in the room than in the plenum. More generally, for other combinations of $\gamma_{0}$ and $\gamma_{1}$, the physical interpretation of the slow and fast manifolds is less straightforward, as they involve other linear combinations of the concentrations in the rooms, which depend on the mixing properties of the system. 

\section{Data analysis and comparison with the model}
\label{sec:model-data}
This section is dedicated to analysing CO$_2$ measurements in the laboratory using the phase space representation, followed by comparison with the predictions obtained from the model presented in \S\ref{sec:theory}. 

Table \ref{tab:input} summarises the input values for the analytical model in \eqref{eq:room} and \eqref{eq:plenum}. The value for the volume flux $q$--the fresh air fed to the plenum--is directly measured by sensors placed in the supply and return ducts with a sampling interval of 5 minutes; table \ref{tab:input} gives the average calculated over a month of measurements. $Q$--the volume flux from the FCU to the occupied room--is not routinely measured, but is obtained by taking ad hoc measurements with an air speed transducer (see \S\ref{sec:volume_flux}).
The forcing term $NF$ is calculated using the occupancy estimation from WiFi connection measurements (for $N$) and taking $F = 0.012$ g/s per person.
\begin{table}
\centering
\begin{tabular}{@{}lcc@{}}
\toprule
 & Occupied space & Plenum \\
\midrule
Volume  & $V_0=$ \SI{512}{\cubic\meter} & $V_1=$\SI{80}{\cubic\meter} \\
Mixing parameter & $0<\gamma_0<1$ & $0<\gamma_1<1$\\
Volume flux (ON) & $Q =$ \SI{0.8}{\cubic\meter\per\second} & $q=$ \SI{0.25}{\cubic\meter\per\second}\\
Volume flux (OFF) & $Q =$ \SI{0.01}{\cubic\meter\per\second} & $q=$ \SI{0.02}{\cubic\meter\per\second}\\
\bottomrule
\end{tabular}
\caption{Input parameters for the analytical model \eqref{eq:room} \eqref{eq:plenum}. The values of $q$ are measured by the BMS system. The forcing $NF$ is estimated by the measured WiFi connections and $F = 0.012$ g/s per person. $Q$ is estimated by air speed measurements done in the room (see appendix \S\ref{sec:inf_exf}).}
\label{tab:input}
\end{table}
%
%
\begin{figure}
\centering
\includegraphics[width=0.9\linewidth]{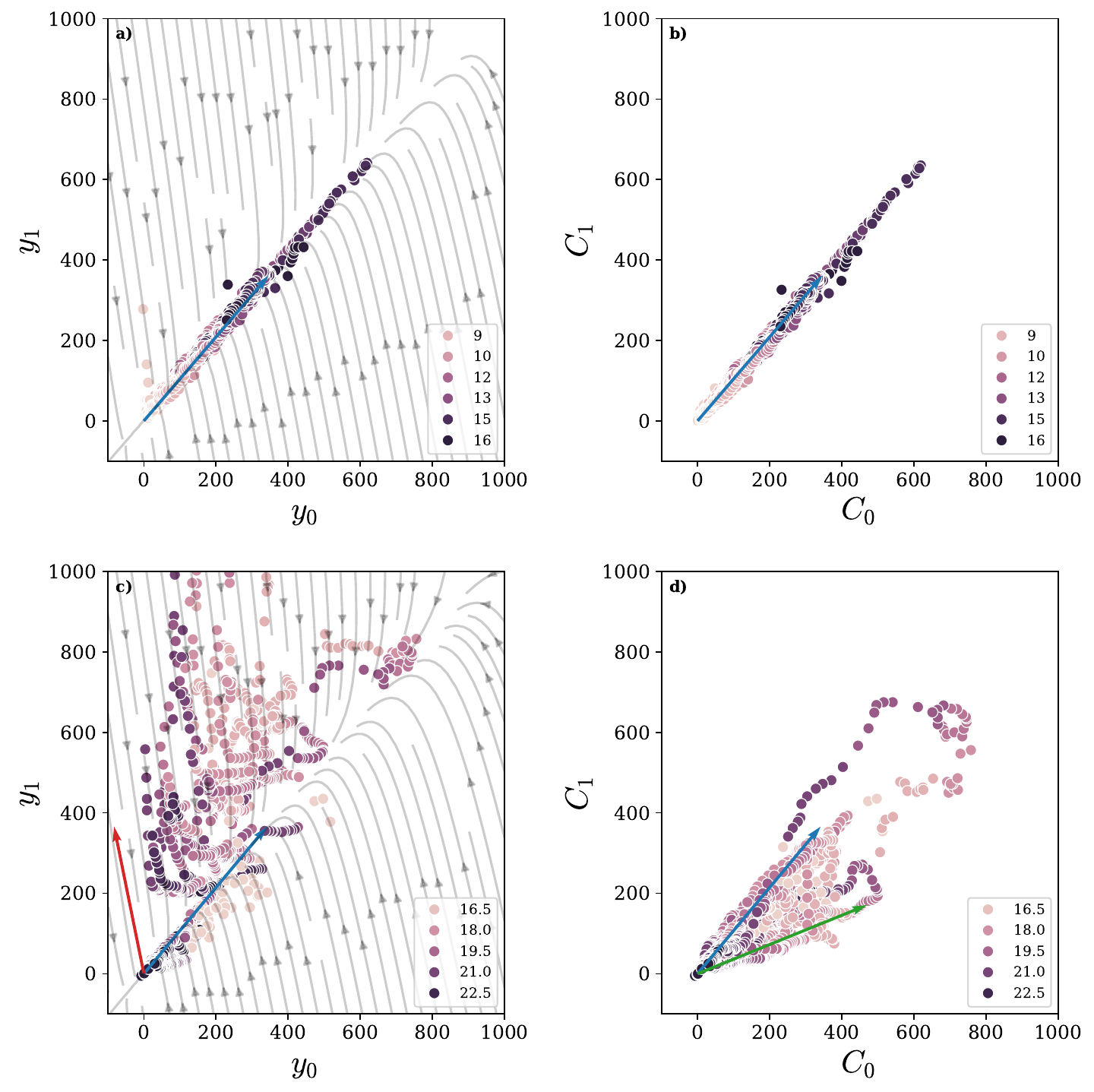}
\caption{Phase space diagram showing the excess of CO$_2$ data in the room ($C_0$) and in the plenum ($C_1$) for July (excluding weekends). The spaces in the $y-$coordinate system in (a) data when ON, (c) data when OFF. The blue arrow marks the eigenvector related to the slow manifold and the red arrow in (c) the fast manifold both calculated for $\gamma_0=1$ and $\gamma_1 = 0$. The gray arrow show the flow evolution trajectories as in figure (\ref{fig_phasetheory}). The same dataset is plotted in the $C_0, C_1$ space in (b) for data when ON and (d) data when OFF. The blue arrow shows the eigenvector calculated for $\gamma_0=1$ and $\gamma_1 = 0$ and the green arrow marks the eigenvector for $\gamma_0=0.2$ and $\gamma_1 = 0.5$. The markers colour represent the time of the day, as indicated in the legend.}
\label{fig_phase_dataJuly}
\end{figure} 

Figure \ref{fig_phase_dataJuly} shows the measured excess of CO$_2$ in the room ($C_0$) and in the plenum ($C_1$) for the entire month of July (weekends excluded) in a phase space diagram. The dataset is divided into two sets corresponding to the ON regime (figure \ref{fig_phase_dataJuly}  (a) and (b)) and the OFF regime (figure \ref{fig_phase_dataJuly}(c) and (d)). Panels (a) and (c) depict the data in the $y-$coordinate system (as per \eqref{eq:y}). The model (in terms of the $\boldsymbol{y}$ coordinate) is also shown on these panels as a reference (with the grey arrows marking the flow evolution trajectories, and the red and blue arrow marking the eigenvectors representing the fast and slow dynamics of the system, respectively). Panels (b) and (d) show the measured data in terms of excess CO$_2$ concentrations. The volume fluxes and volumes are inputted into the model to form the matrix $\boldsymbol{A}$ according to the measured values (see table \ref{tab:input}), while the parameters $\gamma_0$ and  $\gamma_1$ cannot be measured and therefore need to be fitted with the data. For the case ON, the data are always well represented by $\gamma_0=1$ and $\gamma_1=0$, which correspond to strong mixing within the room and to supply air that enters the FCU having effectively bypassed the plenum, respectively. It is striking how the concentrations converge rapidly onto the subspace describing the slow dynamics of the system, marked by the blue arrow. This convergence points to a significant time separation between the slow and fast dynamics of the system. Another characteristic we can deduce from this plot is that the exchange driven by the FCU could be considered as “overdriving" the response of the room in the sense that lowering the fan speed setting would still achieve a quick collapse onto the slow dynamics line.

The data are more scattered in the OFF regime (figure \ref{fig_phase_dataJuly} (c) and (d)), for which $q$ and therefore $Q$ are significantly smaller than in the ON regime (but are not zero, see table \ref{tab:input}). The comparison with the model in panel (c) highlights how some data lie on the slow manifold, but many sequences of points align with the fast manifold indicated by the red arrow. This behaviour reveals an example of a regime where the two-dimensionality of the system becomes important and the timescale on which the forcing varies, relative to the timescale characterising the fast eigenmode, has changed markedly (compared with that in the ON regime) with respect to that of the fast eigenmode response.
Panel (d) shows how the data might be represented by the model with a range of $\gamma$ parameters, with a limiting case that corresponds to the same conditions as in the ON regime (marked by the blue arrow). Most of the other data lies `below' this eigenvector (i.e.\ $C_1 < C_0$) and the model can be fitted using a range of combinations of $\gamma_0$ and $\gamma_1$, suggesting that mixing in both the room and the plenum plays an important role in the OFF regime. In figure \ref{fig_phase_dataJuly} (d), the green arrow marks the eigenvector calculated by setting $\gamma_0=0.2$ and $\gamma_1 = 0.5$. Since there is little sensitivity to $\gamma_1$ for most values of $\gamma_0$, similar eigenvectors can be obtained for different values of $\gamma_1$. 
\begin{figure}[t]
\centering
\includegraphics[width=0.9\linewidth]{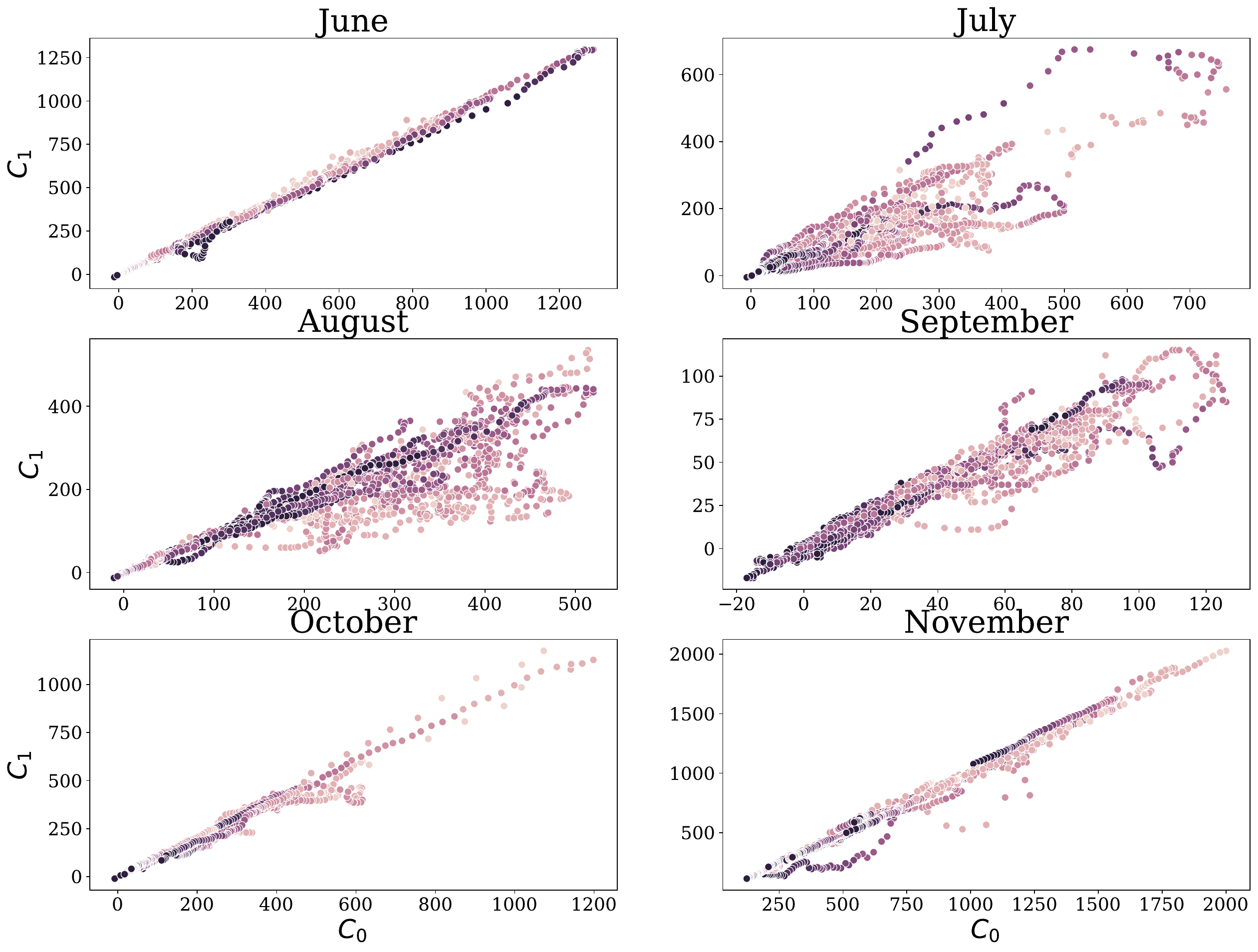}
\caption{Phase space diagram showing the CO$_2$ data in the room ($C_0$) and the plenum ($C_1$) for months between June 2022 and November 2022 at the times when the FCU is off.}
\label{fig_phase_dataMonths}
\end{figure} 

To better understand what causes the spread in the OFF regime, we plotted the phase space diagrams in figure \ref{fig_phase_dataMonths} to compare the CO$_2$ concentrations over six months, from June to November. The plots clearly show that during the warmer months (July, August, and September), the data are more spread, while in the colder months, they align along the eigendirection associated with the slow dynamics.%
\begin{figure}[t]
\centering
\includegraphics[width=0.9\linewidth]{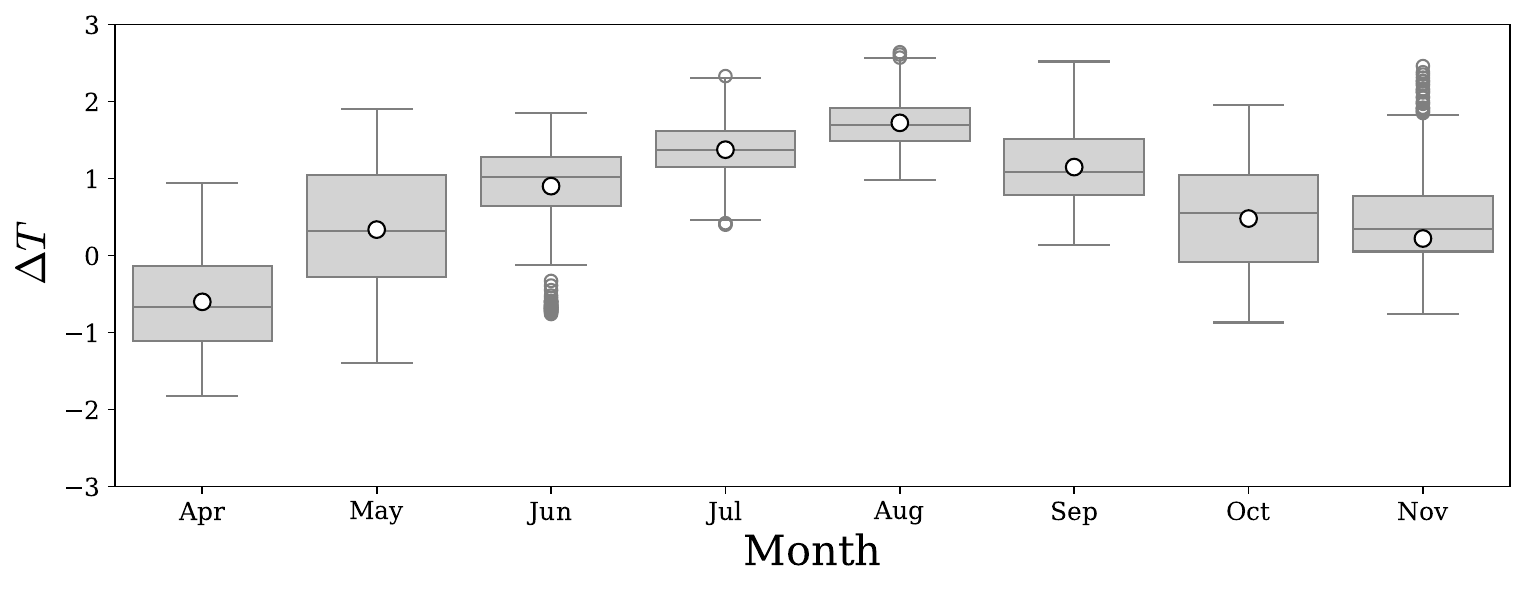}
\caption{Boxplot of the monthly averaged temperature difference between plenum and room. The horizontal line within each box marks the median, and the boxes extend from the first to the third quartile. The white dot marks the mean value. The diamond-shaped markers indicate the outliers.}
\label{fig_TsTr}
\end{figure} 
Given the time of the year for which such differences appear, a reasonable explanation involves a link between the external temperature and the mixing inside the room and plenum. To demonstrate this link, we look at the difference in temperature recorded by BMS in the plenum and in the room, which shows an increase in the temperature from the plenum to the room in the months of July, August, and September (see figure \ref{fig_TsTr}). During the warmer months, the external air supplied to the plenum is warmer than the air in the room. In this situation, the air transferred from the plenum into the room is warmer than that in the room. This lighter and warmer air enters the room through displacement, causing stratification in the vertical direction.
On the contrary, when the air entering the room is colder, this heavier air descends readily, causing stirring and mixing. This analysis highlights that buoyancy-driven flow -- the effects of which are represented by the mixing parameter -- might play an important role in determining the distribution of CO$_{2}$.

%
\begin{figure}
\centering
\includegraphics[width=0.9\linewidth]{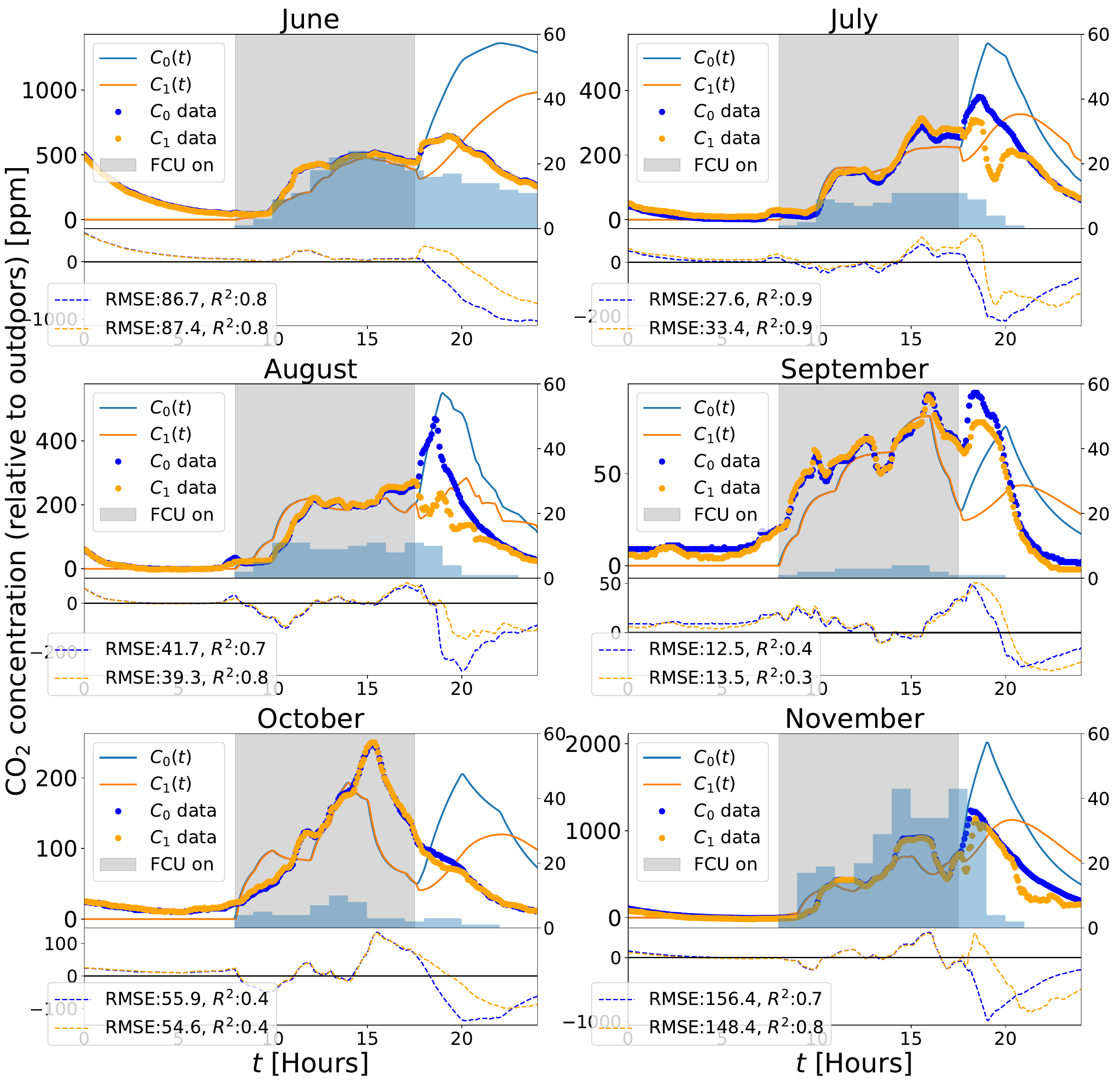}
\caption{CO$_2$ concentration (relative to outdoors) time evolution and model prediction. The data (blue and yellow dots) correspond to the measured concentrations in the laboratory on a representative day of the dataset (the first Tuesday of each month). The ambient CO$_2$ (410 ppm) has been subtracted from the data. Note that the range of CO$_2$ concentrations (left $y-$ axis) differs significantly from month to month. The solid yellow and blue lines represent the solution of \eqref{eq:room_a} and \eqref{eq:plenum_a}; see text for more details. The blue-shadowed histograms indicate occupancy in terms of the number of people in the room. The grey-shadowed region marks the ON regime. The dashed curves under each plot represent the residuals and the RMSE and $R^2$ values for the ON phase are annotated in the caption.}
\label{fig_timedata}
\end{figure} 

Finally, we can use the theoretical model to study the time evolution of the CO$_2$ within the room and the plenum. Figure \ref{fig_timedata} illustrates such a comparison for a chosen day (the second Tuesday for each month from June to November). The dataset spans very different conditions, with the occupation in any given hour ranging from a few people to over 50. The outside air temperature ranges from a few degrees Celsius in the autumn months to above 30 degrees Celsius in the months of July and August.

The data (plotted in blue and yellow dots) corresponds to the CO$_{2}$ concentrations in the laboratory and plenum. The solid blue and yellow lines represent the solution of \eqref{eq:room_a} and \eqref{eq:plenum_a} fed by the experimental values given in table \ref{tab:input}. The mixing parameters are set to $\gamma_0 =1$ and $\gamma_1 = 0$ for the ON case, the latter indicating that the totality of the fresh air goes into the FCU without mixing with the air in the plenum. As discussed previously, the OFF case is more complex. In figure \ref{fig_timedata}, we fixed the parameters to $\gamma_0= 0.2$ and $\gamma_1= 0.5$ for all months for simplicity.

Considering the uncertainty related to the occupancy estimation (see \S\ref{sec:occupancy}) and that occupancy is sampled only hourly whilst the data are sampled every 5 minutes, the model is in good agreement with the measurements. For the regime ON the $R^2$ values are above 0.7, except when occupancy values are below 5 (i.e.\ the CO$_2$ forcing is relatively weak), as in the September and October panels in figure \ref{fig_timedata} for which the $R^2$ values drop below 0.4. The RMSE value plots show that the largest errors correspond with the OFF regime and with periods of large relative variations in occupancy that again are coarsely sampled in time. When the FCUs are on (in the shaded grey region of figure \ref{fig_timedata}), the CO$_2$ concentration in the room and the plenum are nearly always equal. On the contrary, when the FCUs are switched off, the CO$_2$ response to the forcing is more unpredictable, and the concentration difference between the plenum and the room varies.

\section{Model reduction and generalisation to higher dimensions}
\subsection{Model reduction}
\label{sec:modelReduction}
We revisit the model presented in \S\ref{sec:model} in light of the behaviour highlighted by the comparison with the measured data. Numerous situations have emerged in which there is a sufficiently large separation of timescales that, from the perspective of the relatively slow evolution of the overall system, the transient effects associated with the fast dynamics can be neglected. Exploiting a separation of time scales in this way is used widely to derive a simplified low-dimensional description of a physical system and, in broad terms, corresponds to the `adiabatic elimination' of fast variables in stochastic systems \citep{HakHboo1983a}.

The equation governing the second (fast) eigenmode can be written (Appendix \ref{sec:solution}; \eqref{eq:z})
\begin{equation}
    \frac{\rd z_2}{\rd\tau} = \lambda_2 z_2+ \frac{1}{\lambda_2 R_{21}}\frac{\rd f_0}{\rd \tau} ,
    \label{eq:dz2dt}
\end{equation}
where $|1/\lambda_2|$ is the response timescale of the eigenmode, $z_2$ is the amplitude of the eigenmode, and $\bold{R}$ is the eigenvector matrix. Since $\lambda_{2}<0$, we can deduce that an upper bound for finite changes $\Delta z_{2}$ is obtained by ignoring the first term on the right-hand side of \eqref{eq:dz2dt}: $\Delta z_{2}\leq \Delta f_{0}/(\lambda_{2} R_{21})$. The upper bound becomes tighter in the limit that the corresponding $\lambda_{2}\Delta t\rightarrow 0$. For larger values of $\Delta t$ the upper bound becomes increasingly conservative. In other words, it is changes in $f_0$ over timescales shorter than $1/\lambda_2$ that are likely to lead to devations from $z_{2}\approx 0$, which would remain a valid approximation, and therefore enable a reduction of the dimension of the system, if $\Delta f_{0}\ll \lambda_{2}R_{12}$. In practice, this condition might not be fulfilled when there are rapid changes in occupancy, such as at the beginning and end of a class. 

By assuming that the equation for $z_{2}$ plays no role, the (one-dimensional) reduced model can be written (Appendix \ref{sec:solution}; \eqref{eq:z}) as
\begin{equation}
    \frac{\rd z_1}{\rd\tau} = \lambda_1 z_1+ \frac{1}{\lambda_1}R^{-1}_{11}\frac{\rd f_0}{\rd \tau} ,
    \label{eq:reducedZ}
\end{equation}

\noindent and the solution \eqref{eq:Csol} reduces to

\begin{equation}
\boldsymbol{C} = \boldsymbol{R}_{1}\exp(\lambda_1 \tau)\boldsymbol{R}_{1}^{-1}(\boldsymbol{C}_0 +\boldsymbol{A}^{-1}\boldsymbol{f})-\boldsymbol{A}^{-1}\boldsymbol{f},
\label{eq:Csolreduced}
\end{equation}
where $\boldsymbol{R}_{1}$ is the column vector of the eigenvector matrix (corresponding to the two components of one eigenvector), and $\boldsymbol{R}_{1}^{-1}:=\boldsymbol{R}_{1}^{\top}/(\boldsymbol{R}_{1}^{\top}\boldsymbol{R}_{1})$ is the left inverse. Note that \eqref{eq:Csolreduced} depends only on the eigenvalue and eigenvector associated with the slow dynamics.

\begin{figure}[t]
\centering
\includegraphics[width=0.99\linewidth]{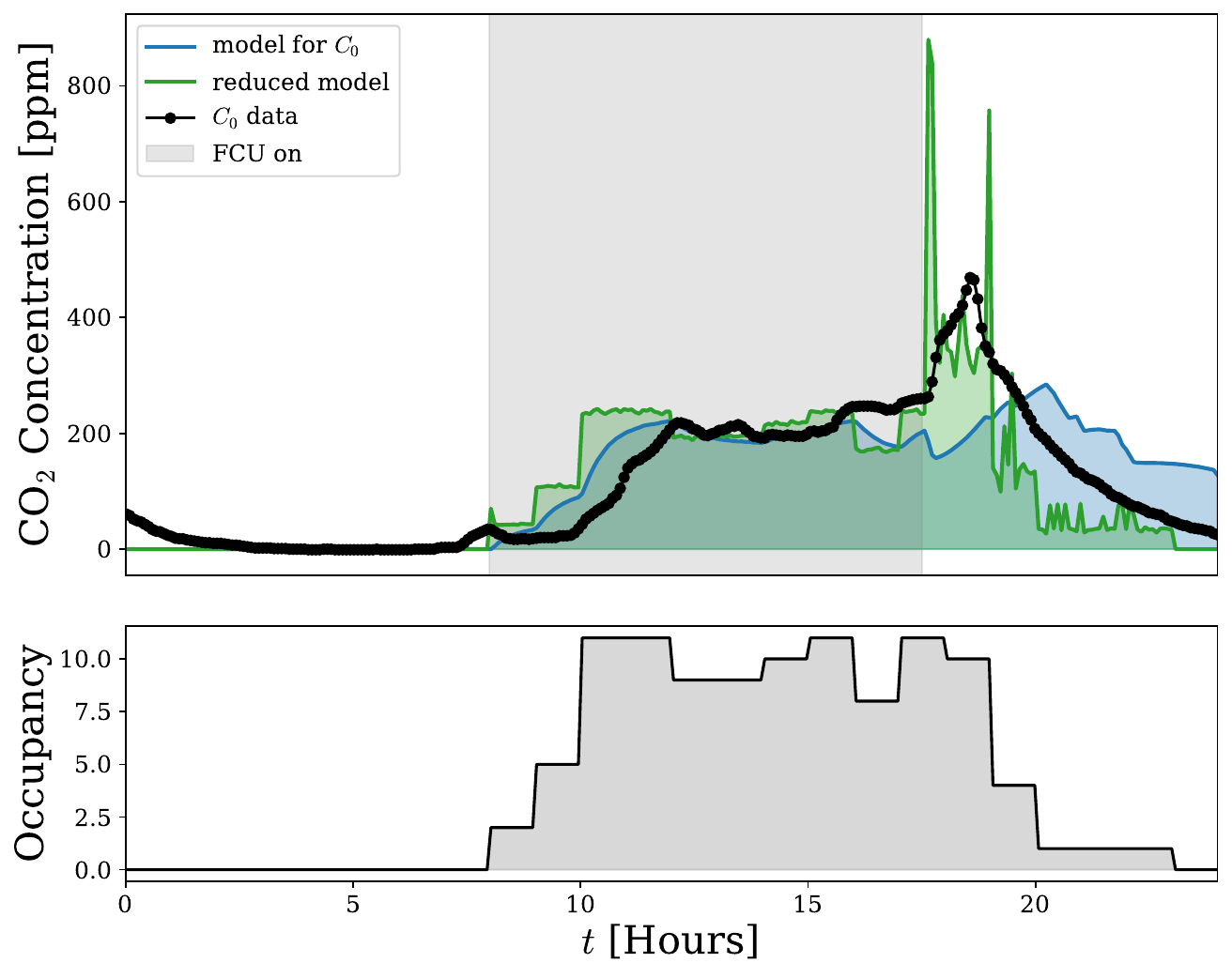}
\caption{Top panel: comparison of the reduced (green line) and full model (blue line) with the observational data (black dots) for the month of August. The parameters for the two models are the same as the one used in figure \ref{fig_phase_dataMonths}. Bottom panel: occupancy data.}
\label{fig_reducedmodel}
\end{figure} 
The full and reduced models are compared in the top panel of figure \ref{fig_reducedmodel} (blue and green line, respectively), where the time evolution of $C_0$ is plotted for a chosen day in August (same data as in figure \ref{fig_phase_dataMonths}). The reduced model prediction (given by \eqref{eq:Csolreduced}) is consistent with the full model prediction (given by the integration of \eqref{eq:room}). We can notice that the rapid changes in occupancy is approximated in the reduced order system by an instantaneous adjustment of the fast mode and are responsible for the differences between the outputs from the full and reduced order models (top panel).

\subsection{Generalisation to larger dimensional models}
\label{sec:modelMoredim}
The model presented in \S\ref{sec:model} can be generalised to analyse the dynamics of any arbitrary $n-$zone configuration. 
The system is expressed by \eqref{eq:x}, in which $\bold{C}$ is now a vector of size $n$, and  $C_i$ represents the CO$_2$ concentration in the $i$th zone. $\bold{A}$ becomes a $n \times n$ matrix, the elements $A_{ij}$ of which express how the CO$_2$ transport occurs between zones $i$ and $j$. In general, $\bold{A}$ cannot be diagonalised, but we can use the singular value decomposition (SVD), a matrix factorisation technique that is widely used to generalise the eigendecomposition \cite{brunton2022data}. The matrix $\bold{A}$ can always be decomposed into $\bold{A}=\bold{U}\bold{\Sigma}\bold{V^*}$, where $\bold{U}$ and $\bold{V}$ are unitary matrices with orthonormal columns ($^*$ indicates the complex conjugate transpose), and $\bold{\Sigma}$ is a diagonal matrix of non-negative real numbers. The columns of $\bold{U}$ and the columns of $\bold{V}$ are called the left-singular vectors and right-singular vectors of $\bold{A}$, respectively, and they form two sets of orthonormal bases. The diagonal elements of $\bold{\Sigma}$ are called singular values, and it is always possible to choose a decomposition such that the singular values are in descending order. 

Analogously to what we have done for the two-room (four-zone) system, analysing the singular vectors and singular values allows identification of both the important parameters driving the system and those that can be considered redundant. This analysis can therefore lead to a simplified representation of the multizone space, with the number of singular values retained in a reduced-order model being dictated by the given application. Here, we have studied the evolution from initial conditions in highly transient problems and the timescale analysis has proven successful in isolating slow and fast modes that will and will not respond strongly to the forcing.

\section{Discussion}
\label{sec:discussion}
An analytical multi-zone ventilation flow model and a dynamical systems analysis have been applied in a  four-zone case study to demonstrate the potential of the approach in detail. We have further reasoned that the analysis can be extended to any ventilated multi-zone space, as discussed in \S\ref{sec:modelMoredim}. It is important to note that although our case study focuses on a mechanically ventilated space, the same methods can be applied to any HVAC control system and even to naturally ventilated buildings, provided the airflows are known.

The analysis of the timescales involved in the system helps identify a slow manifold to capture the bulk response behaviour, dimensionality reductions that can be made in modelling the system and the transient regimes that play a significant role during adjustment processes.
For instance, the data analysis in \S\ref{sec:model-data} highlights the rapid convergence in the ON regime of CO$_2$ concentration onto the slow dynamics of the system. A dimensionality reduction of the system response is possible, i.e. a collapse onto a single dimension in this instance, because the resulting timescale separation between the fast and slow dynamics is more than a factor of 10: the characteristic fast timescale is around 30 seconds, while the slow timescale exceeds 5 minutes (characteristic timescales are calculated as $1/\lambda$ and then dimensionalised as per \eqref{eq:adim}). 

More generally, we can analyse the eigenvalues expressed in \eqref{eq:eigenvaluesExplicit}, which correspond to the magnitude of the associated eigenvectors, to explore the sensitivity of the system in parameter space.
\begin{figure}[t]
\centering
\includegraphics[width=0.99\linewidth]{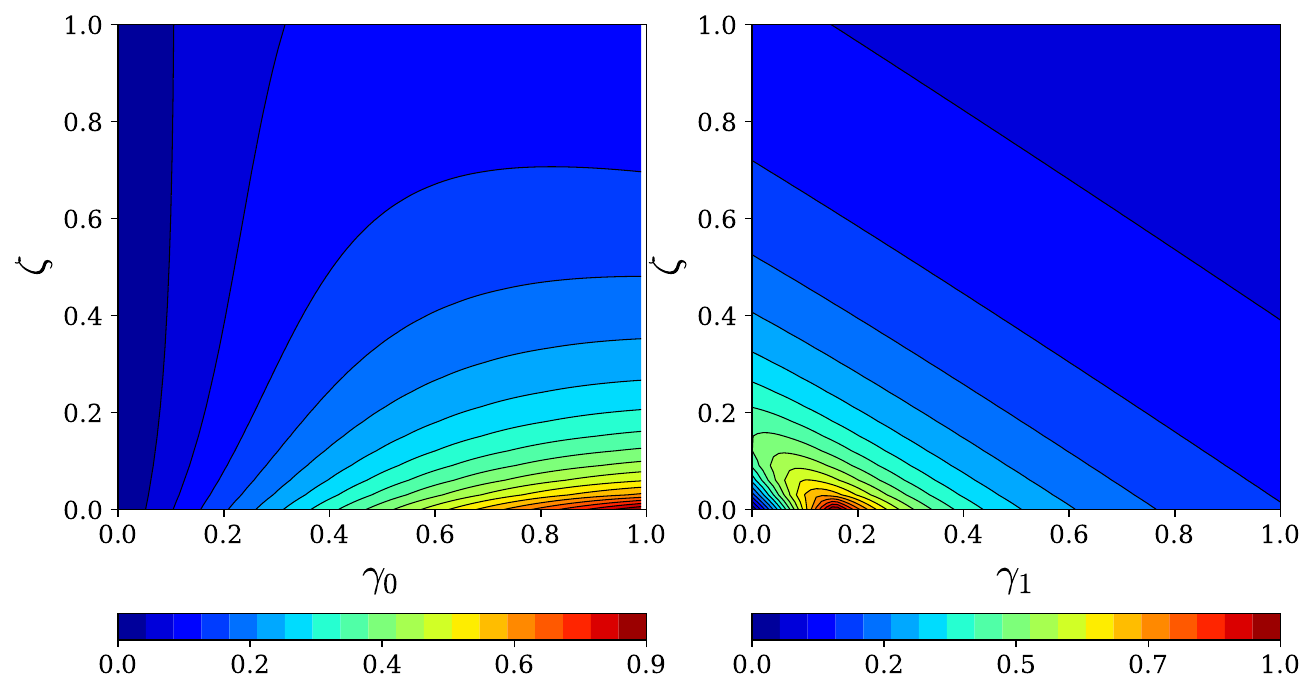}
\caption{Ratio of eigenvalues (indicated by colour)  as a function of the dependent variables (left) $\zeta$ and $\gamma_0$ with $\gamma_1 = 0.18$, and (right) $\zeta$ and $\gamma_1$ with $\gamma_0=1$. $\varepsilon = 0.15$ for both panels.}
\label{fig_dimensionreduction}
\end{figure} 
For $\varepsilon \ll 1$ in our case study, the two eigenvalues are of a different order of magnitude throughout most of the parameter space, supporting characterisation by a one-dimensional dynamical system. Regions where this dimensionality reduction becomes invalid can also be found along with the dependence on the values of the other three variables: $\zeta$, $\gamma_0$, and $\gamma_1$. Figure \ref{fig_dimensionreduction} illustrates such dependencies for $\varepsilon = 0.15$, with the colours corresponding to the ratio $\max (\lambda_1, \lambda_2)/(\min(\lambda_1, \lambda_2))$. The reduction to a one-dimensional description is expected to hold throughout most of the parameter space because values of the ratio of the eigenvalues are close to zero, consistent with the finding (by inspection)  in previous sections. If fresh air is introduced in the system, even a surprisingly small recirculation $q' \gtrapprox 0.3 q$ is sufficient to promote collapse onto the one-dimensional system. However, for $\zeta \lessapprox 0.3$, there are always combinations of $\gamma_0$ and $\gamma_1$ for which the system remains bi-dimensional.

In addition to suggesting a leading order representation of complex ventilation systems, our approach helps to identify when and how a system is running inefficiently and therefore how the running cost and energy requirements might be reduced. The total energy consumption of an HVAC system depends on numerous variables, including the supply air flow rate, and is typically complex to estimate  \cite{atthajariyakul2004real}. Nevertheless, in the case study presented here, our analysis indicates that for optimal CO$_2$ concentration distribution between the room and the plenum, the air recirculated by the FCUs need generate a recirculation that is only 30\% of the fresh air supplied into the plenum. From this perspective, the excess air recirculated to the room is overdriving the system without providing any real benefit in pollutant removal. The model suggests that the FCU air recirculation for the studied configuration could be reduced by more than 50\%, resulting in a significant reduction in costs without impacting the air quality experienced by occupants. 

To estimate the potential energy saving, let us consider the annual energy usage of the FCUs in the room. According to the technical specifications, each FCU installed in the room has a fan operation electrical power consumption of \SI{0.3}{\kilo\watt} for the high speed setting \cite{webSapphire}. Considering that the FCUs are used for an average of 8 hours per day and that there are 4 FCUs installed in the room, the estimated annual energy consumption is approximately \SI{3500}{\kilo\watt\hour}/year. Note that this estimate focuses on fan power only because the thermal condition of the room still needs to be maintained. 
If we halve the FCU flow rate as suggested by the model, this will result in a decrease in annual energy usage by \SI{1750}{\kilo\watt\hour}/year. In terms of energy consumption per unit area, this reduction corresponds to \SI{10.9}{\kilo\watt\hour\per\square\meter}/year (for a room with a surface area of \SI{160}{\square\meter}), consistent with UK ``good practice''." - e.g. \href{https://bregroup.com/documents/d/bre-group/d_airconditioningenergyuseannexd}{bregroup}.

\section{Conclusions}
\label{sec:conclusions}
We have studied the evolution of pollutant concentration in connected spaces by combining analytical modelling with analysis of observational data from a living laboratory. The use of the data to validate the modelling has offered a robust approach for predicting pollutant concentrations in a ventilated space. 

The formulation of an analytical model depends upon first identifying a number of  zones in the ventilated space along with key parameters describing the zone geometries, the flow between zones and the mixing within each zone. The methodology then utilises an eigenmode decomposition to analyse the model properties in phase space. The aim is finally to identify an appropriate  reduced-dimension description of the ventilation flow system. 

In the living laboratory examined in this study, we proposed several zones (sub-volumes) to represent the potential heterogeneity in the pollutant (CO$_2$) distribution. The analysis highlighted the importance of parameters describing the relative volumes of the selected zones; in particular, significantly different time scales associated with the ventilation of each zone  proved to be a key step towards a significantly simplified model. 
Moreover, although the values of some other parameters in the model were not known, the approach allowed the sensitivity of the system to those parameters to be quantified. Throughout the parameter space considered in this study, we were thus able to show that detailed knowledge of these parameters was unimportant.

Ventilation flows and pollutant distribution in buildings are undoubtably governed by more complex physics than we have attempted to represent, but the present study has highlighted some useful general principles. Firstly, and somewhat ironically, a good operational balance between ventilation and energy consumption seems likely when an inadequate model would result from a reduced-order representation, i.e.\ the eigenvalues of a modal decomposition have comparable values. Secondly, it is apparent that knowledge of pollutant sources -- in this case, spatio-temporal variability in occupancy -- represents a major source of uncertainty for models, in general. These effects are typically overlooked in  modelling approaches that assume spaces to be well-mixed. However, the dataset available from the numerous sensors situated throughout our living laboratory has the potential to help answer outstanding questions about spatial pollutant distribution, but a key future step is to better account for  spatio-temporal characteristics (including any associated thermal effects) of the corresponding sources.
Better understanding of potential accumulation zones for pollutants would help improve design and placement of air extractors and benefit air quality and safety.


\paragraph{Acknowledgements}
We are grateful to Mark Reader for providing the BMS data and insightful comments about the HVAC system. We thank Neal Streamer and Trend Controls for supporting the project and providing the sensors and controllers. We thank Chris Banks from Imperial College London for facilitating access to the occupancy data from Imperial's deployment of the HubStar (formerly LoneRooftop) Building Insights Dashboard which infers occupancy from WiFi connections.

\paragraph{Funding Statement}
This work was supported by the Engineering and Physical Sciences Research Council [grant number EP/V033883/1]
as part of the [D$^{*}$]stratify project.

\paragraph{Ethical Standards}
The research meets all ethical guidelines, including
adherence to the legal requirements of the study country.

\paragraph{Supplementary Material}
\bibliographystyle{elsarticle-num} 
\bibliography{biblio1}
\newpage
\appendix
\section{Supplementary material 1: analytical model and phase space}
\label{sec:solution}
\subsection{Solution of the CO$_2$ model}
The following equation describes the plenum-room system
\begin{equation}
 \frac{\rd\boldsymbol{C}}{\rd\tau} = \boldsymbol{A}\boldsymbol{C}+\boldsymbol{f}
\label{eq:xA}
\end{equation}
where $\boldsymbol{A}$ is a coefficient matrix, $\boldsymbol{C} = (C_0, C_1)^{\top}$ is a state vector and $\boldsymbol{f}=(f_0, 0)^{\top}$ is a forcing vector. Equation \eqref{eq:xA} can be converted into an autonomous form by subtracting the steady-state solution from $\boldsymbol{C}$:

\begin{equation}
\boldsymbol{y} := \boldsymbol{C}+\boldsymbol{A}^{-1}\boldsymbol{f},
\label{eq:suby}
\end{equation}
so that
\begin{equation}
 \frac{\rd\boldsymbol{y}}{\rd\tau} = \boldsymbol{A}\boldsymbol{y}+\boldsymbol{A}^{-1}\frac{d\boldsymbol{f}}{d\tau},
\label{eq:yA}
\end{equation}
where the second term on the right-hand side accounts for temporal changes in the forcing. Assuming that $\boldsymbol{A}$ is diagonalisable, let us now factorise $\boldsymbol{A}$: 

\begin{equation}
\boldsymbol{A} = \boldsymbol{R} \boldsymbol{\Lambda}\boldsymbol{ R}^{-1},
\label{eq:basis}
\end{equation}
where the eigenvalue matrix is
\begin{equation}
\boldsymbol{\Lambda} = \begin{bmatrix} \lambda_1 & 0\\0 & \lambda_2 \end{bmatrix} 
\end{equation}
and $\boldsymbol{R}$ is the corresponding (right) eigenvector matrix
\begin{equation}
\boldsymbol{R}= \begin{bmatrix} v_1^{(1)} & v_1^{(2)}\\v_2^{(1)} & v_2^{(2)} \end{bmatrix} 
\end{equation}
with the columns corresponding to each eigenvector $\bold{v_1} = (v_1^{(1)}, v_2^{(2)})$, $\bold{v_2} = (v_1^{(1)}, v_2^{(2)})$.

By substituting (\ref{eq:basis}) into (\ref{eq:yA}), premultiplying by $\boldsymbol{R}^{-1}$ and defining 
\begin{equation}
\boldsymbol{z} := \boldsymbol{R}^{-1}\boldsymbol{y},
\label{eq:subz}
\end{equation}
equation \eqref{eq:yA} can be expressed in the simplified form
\begin{equation}
\frac{\rd\boldsymbol{z}}{\rd\tau} = \boldsymbol{\Lambda}\boldsymbol{z}+ \boldsymbol{R}^{-1}\boldsymbol{A}^{-1}\frac{\rd\boldsymbol{f}}{\rd\tau} ,
\label{eq:z}
\end{equation}
where $\boldsymbol{z}$ represents the amplitude of each eigenmode. If the forcing term $\boldsymbol{f}$ is independent of time, then the system becomes autonomous:
\begin{equation}
\frac{\rd\boldsymbol{z}}{\rd\tau} = \boldsymbol{\Lambda}\boldsymbol{z},
\label{eq:zs}
\end{equation}
and has the solution
\begin{equation}
\boldsymbol{z} = \boldsymbol{z}_{0}\exp(\boldsymbol{\Lambda} \tau),
\end{equation}
or, transforming back to the original coordinate system,
\begin{equation}
\boldsymbol{C} = \boldsymbol{R}\exp(\boldsymbol{\Lambda} \tau)\boldsymbol{R}^{-1}(\boldsymbol{C}_0 +\boldsymbol{A}^{-1}\boldsymbol{f})-\boldsymbol{A}^{-1}\boldsymbol{f}.
\label{eq:CsolA}
\end{equation}
\subsection{Eigenvalues and eigenvectors}
\label{sec:eigenvectors}
Let us consider a homogenous differential equation for which we want to calculate the eigenvalues $\lambda$ and eigenvectors $\bold{v}$

\begin{equation}
A \bold{v} = \lambda \bold{v}
\end{equation}

We can rewrite the system of non-autonomous linear ODE of the first order in matrix form.

\begin{equation}
\frac{\rd}{\rd t}  \begin{bmatrix} C_0\\C_1 \end{bmatrix} = \underbrace{\begin{bmatrix} \frac{- \gamma_0(q-Q)-\varepsilon}{V_0} & \frac{(q+\gamma_1 Q) + (1-\gamma_0)(q+\gamma_1 Q)}{V_0}\\ \frac{\gamma_0 q}{V_1} & \frac{(q \gamma_0 + Q)(q+\gamma_1 Q)}{V_1(q+Q)} \end{bmatrix}}_{A} \begin{bmatrix} C_0\\C_1 \end{bmatrix} + \underbrace{\begin{bmatrix} NF\\0 \end{bmatrix}}_{K}.
\label{eq:matricC}
\end{equation}

We solve the homogenous differential equation by calculating the eigenvalues $\lambda$ and eigenvectors $\bold{v}$ such that

\begin{equation}
A \bold{v} = \lambda \bold{v}
\end{equation}

If we now look for a solution in the form
\begin{equation}
\begin{bmatrix} C_0\\C_1 \end{bmatrix} =A_0 \begin{bmatrix} v_1^1\\v_1^2 \end{bmatrix} \exp{(\lambda_1 t)} + B_0 \begin{bmatrix} v_2^1\\ v_2^2 \end{bmatrix} \exp{(\lambda_2 t)}.
\end{equation}

For $q'\geq 0$, the eigenvalues are:
\begin{equation}
    \lambda_{1,2} = \frac{1}{2(\zeta+1)} \left [- \varepsilon\gamma_0(1+\zeta)^2 - (\gamma_1 +\zeta)(\gamma_0 \zeta +1) \pm S \right],
    \label{eq:eigenvaluesExplicit}
\end{equation}
with
\begin{dmath}
    S = \left( -4 \varepsilon \gamma_0(\zeta+1)^2(\gamma_1 + \zeta) + ((\gamma_1+\zeta)(\gamma_0 \zeta + 1)+\varepsilon\gamma_0(\zeta+1)^2)^2 \right )^{\frac{1}{2}}
\end{dmath}
The eigenvalues are real, negative, and distinct.
The eigenvectors are

\begin{equation}
    \mathbf{v_1} = \left[- \left (\varepsilon\gamma_0(\zeta+1)^2-(\gamma_1+\zeta)(1+\gamma_0 \zeta) + S_1 \right), 2 \gamma_0 (\zeta+\zeta^2) \right],
\end{equation}
and
\begin{equation}
    \mathbf{v_2} = \left[- \left (\varepsilon\gamma_0(\zeta +1)^2-(\gamma_1+\zeta)(1+\gamma_0\zeta) - S_1 \right), 2 \gamma_0 (\zeta+\zeta^2) \right],
\end{equation}
with
\begin{dmath}
S_1 = \left(\varepsilon^2 \gamma_0^2(\zeta +1)^4 + \varepsilon 2 \gamma_0(\zeta+\gamma_1)(\zeta+1)^2(\gamma_0\zeta-1) + (\gamma_1 + \zeta)^2(1+\gamma_0 \zeta)^2 \right)^{\frac{1}{2}}
\end{dmath}
The span of the two eigenvectors identifies a stable manifold.
\subsection{Phase spaces}
\begin{figure}[t]
\centering
\includegraphics[width=0.99\linewidth]{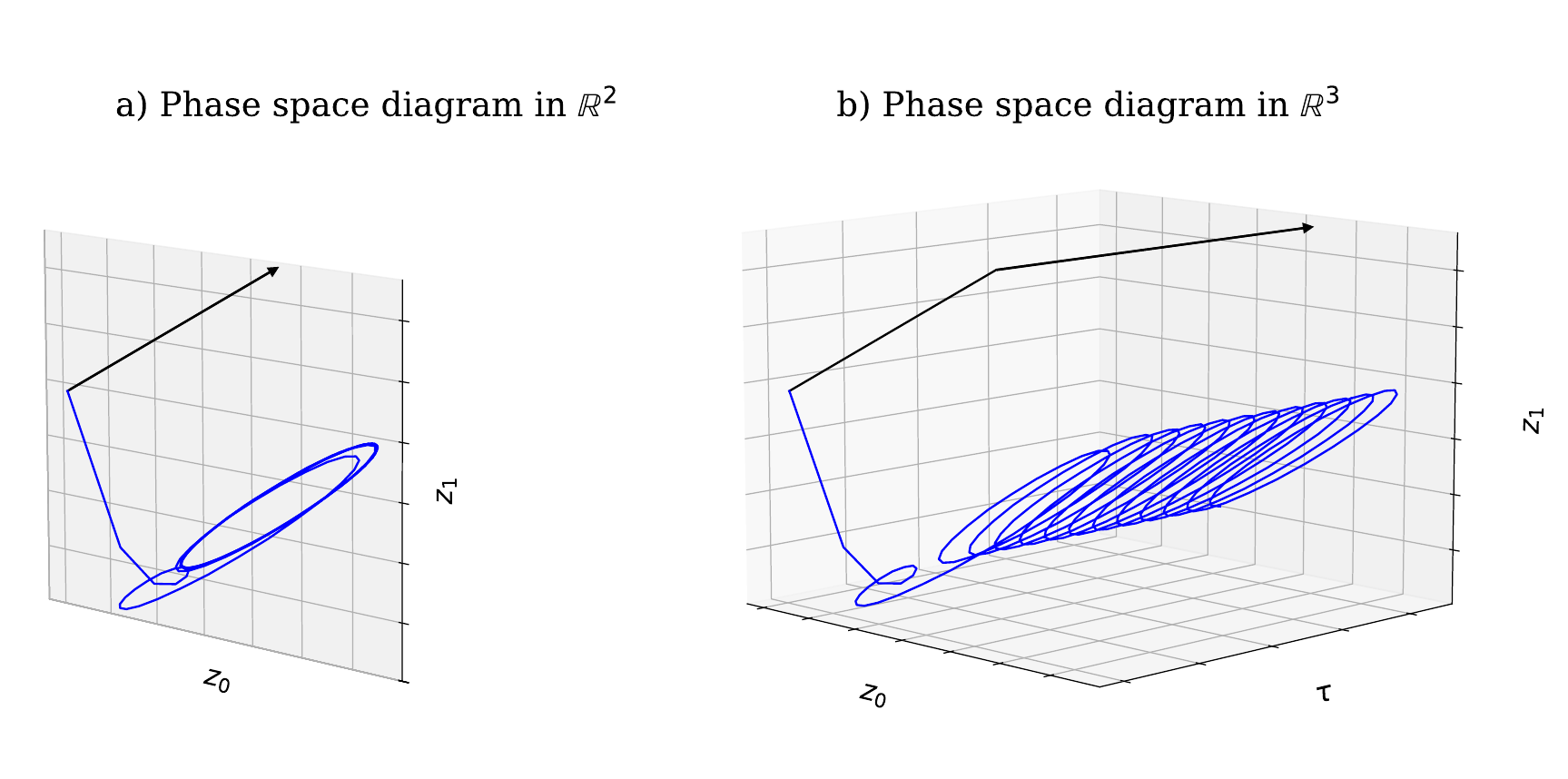}
\caption{Example of phase space diagrams in $\mathbb{R}^{2}$ (a) and in $\mathbb{R}^{3}$ (b) for a forcing that is constant or periodic in time (black and blue line, respectively).}
\label{fig_phaseForcing}
\end{figure} 
For the autonomous system \eqref{eq:zs}, the state
evolution takes place on the phase space $\mathbb{R}^{2}\ni \boldsymbol{z}$ (see the plot in figure \ref{fig_phaseForcing} (a)). Arrows in phase space therefore point in the direction in which $\boldsymbol{z}$ changes with
respect to time, according to \eqref{eq:zs}. Trajectory lines of such a system join up the arrows and cannot cross each other because, for each $\boldsymbol{z}$, \eqref{eq:z} defines a unique arrow direction.

For non-autonomous systems such as \eqref{eq:z}, phase space must be enlarged by one dimension (in such cases, time $\tau$ is
usually introduced as an additional `dependent'/state variable)
to account for the fact that the two-dimensional phase portrait
is affected by time-dependent forcing. Indeed, the
trajectories corresponding to \eqref{eq:z} would, in general,
cross each other in $\mathbb{R}^{2}$ because the right-hand side
of \eqref{eq:z} depends on time. In $\mathbb{R}^{3}\ni
(\boldsymbol{z},t)$, however, trajectories corresponding to
\eqref{eq:z} do not cross (see figure \ref{fig_phaseForcing} (b)). In this work, it will be convenient to plot states and trajectories with respect to the
phase portrait corresponding to \eqref{eq:zs}, acknowledging 
the possibility that the system's trajectories may not be tangential to the arrows when $d \boldsymbol{f}/d\tau \neq \boldsymbol{0}$.
\clearpage
\section{Supplementary material about the living laboratory}
%
\subsection{CO$_2$ concentration vs occupancy estimation}
\label{sec:uncertainty}
\begin{figure}[h!]
\centering
\includegraphics[width=0.99\linewidth]{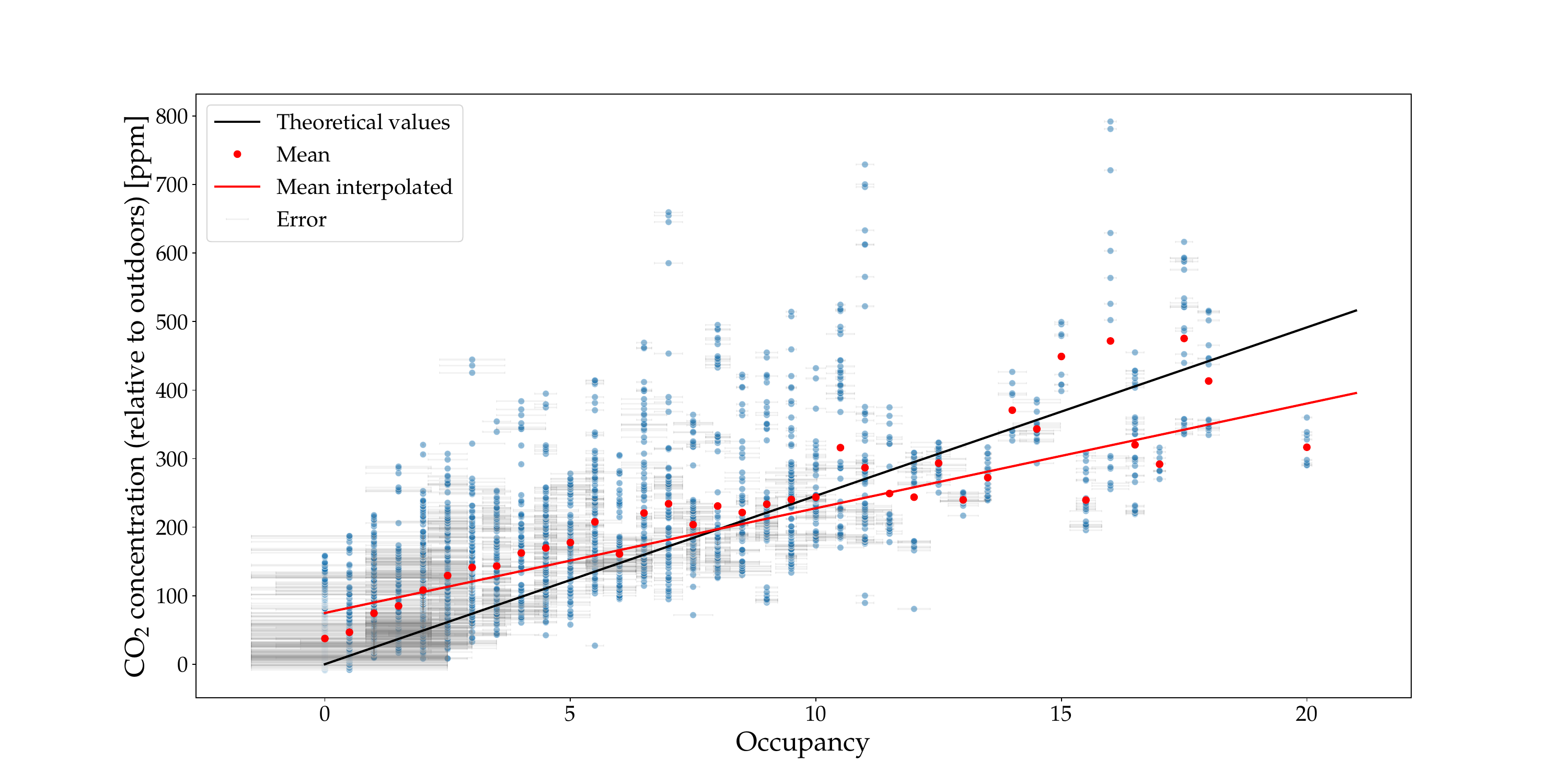}
\caption{Excess CO$_2$ concentration (relative to outdoors) measurements from the sensors in the room versus the occupancy from data recorded in July 2022. The black straight line indicates the theoretical values of CO$_2$ concentration from the steady state of the theoretical model. The red dots and line represent the measured data's mean and linear fit.}
\label{fig_occ_CO2}
\end{figure}

Figure \ref{fig_occ_CO2} shows an example of the excess (relative to the outdoor) of CO$_2$ concentration as a function of occupancy for July 2022. The CO$_2$ concentration is based on an average of the 7 sensors placed in the room during this month, and the recorded room occupancy is based on the WiFi proxy. It can be seen that the CO$_2$ concentration tends to increase with the occupancy, although a large spread is evident in the data. The solid black line indicates the theoretical values of CO$_2$ concentration from the steady state of the theoretical model presented in \S\ref{sec:theory}. The gray shaded horizontal bars show the relative error calculated as the difference between the mean and maximum number of connections recorded within one hour divided by the mean connections number.

%

\subsection{Infiltrations and exfiltrations}
\label{sec:inf_exf}

To verify that the supply air flux is equal to the return air flow rate for the laboratory, we consider $q_{\text{in}}$ and $q_{\text{out}}$ data. Their mean values are given in table \ref{tab1} for two regimes: when the FCU is in operation (ON) and when it is not (OFF). The FCU operates on weekdays between 8 am and 5 pm and is switched off during the remaining times. It is worth noticing that when the FCU is not in operation there is still a residual airflow (see \ref{tab1}). During OFF, we can easily see that $q_{\text{in}} \approx q_{\text{out}}$. Hence, the approximation holds. 
During ON, the difference between supplied and return air flux is $\Delta q = q_{\text{in}} - q_{\text{out}} = 13 $ \SI{}{\liter\per\second}. Therefore, the room is slightly pressurised by the excess inflow. The error associated with the assumption $q_{\text{in}} = q_{\text{out}}$ is $5\%$, which we can consider negligible for the purposes of this paper. To account for the difference, we verify whether the flux through the doors is consistent with the estimated leakage. Although the air volumes transported through open doorways can be large \cite{jha2021contaminant}, the total time for which the doors are open is negligible (the doors are alarmed so that they also cannot be kept open for more than 10 seconds). Therefore, we assume that the error associated with neglecting the open door flux is of the order of the measurement error (5 \si{\liter\per\second} as in table \ref{tab1}). For closed doors, we measured the airspeed through the gaps surrounding the two doors, which is $v_{\text{door}}=\SI{0.44}{\meter\per\second}$  Considering the dimensions of the door frame and the few millimetres of space left between the door and the frame, we can estimate a volume flux $\delta_{\text{door}} \approx \SI{11}{\litre\per\second}$. Hence, the door leakage is enough to account for the slight imbalance between the supplied and return flux. Furthermore, the windows in the room are locked and cannot be opened by the occupants. Therefore, we can assume that there are no significant air exchanges other than the ones through the ducts.\\

\subsection{Volume flux between the plenum and the room}
\label{sec:volume_flux}

We first analyse the data provided by the BMS and additionally consider air speed measurements taken in the room to estimate the air fluxes in the laboratory ($q$, $q'$, and $Q$) that are inputs for the model.
The volume flux exchanges among the spaces in the laboratory can be schematically represented by the diagram in figure \ref{fig_sketchSide} (b). 

\begin{table}
\centering
\begin{tabular}{@{}lccc@{}}
\toprule
\multicolumn{2}{c}{BMS data} & ON & OFF \\
\midrule
Fresh air supply flow rate & $q_{\text{in}}$ & $249 \pm 5$ \si{\liter\per\second} & $7 \pm 2$ \si{\liter\per\second} \\
Return air flow rate&  $q_{\text{out}}$ &$ 236 \pm 5$ \si{\liter\per\second} &$ 9 \pm 1$ \si{\liter\per\second}\\
\bottomrule
\end{tabular}
\caption{Volume flux measurements from BMS. The mean and standard deviations for the volume fluxes are calculated over a month of measurements. ON and OFF refer to the FCU being in operation or not, respectively.}
\label{tab1}
\end{table}

The rate at which the FCU pumps fresh air into the room $q$ is obtained from flow rate sensors placed in the supply and the return ducts (giving $q_{\text{in}}$ and $q_{\text{out}}$, respectively; see values in table \ref{tab1}). Infiltration and exfiltration rates were further assessed to be negligible; hence the supply air flux is equal to the return air flow rate. \\

In addition to the volume fluxes provided by the BMS, we assess the recirculation volume flux $q'$ and the FCU volume flux $Q$ that are needed to solve the analytical system. 
Air exchanges between the upper room and the plenum mainly occur through two ceiling grilles and the lateral linear slot diffusers. Minor leaks of air might also occur through other small openings in the ceiling, but we assume they are negligible compared to the flows through the main grilles and slot diffusers. We measured the air speed (with a TSI 8455-03 air velocity transducer, range 0.127 to 50.8 m/s with accuracy $\pm 2\%$) at these main air exchange locations and estimate the flux by multiplying the speed by the corresponding effective area through which the flow occurs. The flux estimates are: $q_{Li} =$ \SI{374}{\litre\per\second} at the lateral linear slot diffusers, and $q_{C1}=$ \SI{46}{\liter\per\second}, and $q_{C2}=$ \SI{113}{\liter\per\second} through the ceiling grilles.
These sum to give
\begin{equation}
    q' = q_{C1} + q_{C2} + q_{Li} \approx \SI{545}{\litre\per\second}.
\end{equation}
If we consider the air speed measured at the linear slot diffuser outlet, we estimate the flux from all four FCUs to be $q_{Lo} = $  \SI{800}{\litre\per\second}. 
We can then verify that
\begin{equation}
Q=q+q' \approx 250 + 545  = \SI{795}{\litre\per\second}  \label{eq:FCU}
\end{equation}
is compatible with the estimate of $q_{Lo}$.
\begin{table}
\centering
\begin{tabular}{@{}lcc@{}}
\toprule
\multicolumn{3}{c}{Opening areas}\\
\midrule
Slot diffuser effective surface area: outlet & $A_{Lo}$ & $0.8$ m$^2$\\
Lateral linear slot diffuser effective surface area: inlet & $A_{Li}$ & $1.7$ m$^2$\\
Louvred face ceiling diffuser effective surface area & $A_{C1}$ & $0.14$ m$^2$ \\
Grille ceiling diffuser effective surface area & $A_{C2}$ & $0.27$ m$^2$ \\
\midrule
\multicolumn{3}{c}{Average air speed}\\
\midrule
Slot diffuser outlet & $v_{Lo}$ & \SI{1}{\meter\per\second}\\
Lateral linear slots inlet & $v_{Li}$ & \SI{0.22}{\meter\per\second}\\
Louvred face ceiling & $v_{C1} $ & \SI{0.33}{\meter\per\second}\\
Grille ceiling & $v_{C2} $ & \SI{0.42}{\meter\per\second}\\
\bottomrule
\end{tabular}
\caption{Geometrical parameters of the laboratory and volume fluxes measurements from BMS. The mean and standard deviations for the volume fluxes are calculated over a month of measurements. ON and OFF refer to the FCU being in operation or not, respectively.}
\label{tab1A}
\end{table}


\end{document}

%% file: diag_v2.tex
\definecolor{cebebeb}{RGB}{235,235,235}
\definecolor{whitesmoke}{RGB}{245,245,245}
\definecolor{gainsboro}{RGB}{220,220,220}
\definecolor{c191966}{RGB}{25,25,102}
\definecolor{cecdf00}{RGB}{236,223,0}
\definecolor{cd24444}{RGB}{210,68,68}
\definecolor{c3a46a3}{RGB}{58,70,163}
\definecolor{c9da5dd}{RGB}{157,165,221}
\definecolor{ca82929}{RGB}{168,41,41}
\definecolor{cc2c2db}{RGB}{194,194,219}

\def \globalscale {1.000000}
\begin{tikzpicture}[y=1cm, x=1cm, yscale=\globalscale,xscale=\globalscale, every node/.append style={scale=\globalscale}, inner sep=0pt, outer sep=0pt]
  \path[fill=cebebeb,line cap=butt,line join=miter,line width=0.0231cm] (15.3336, 23.3218) -- (15.5314, 23.1882) -- (15.5223, 22.4886) -- (15.3336, 22.3796) -- cycle;

  \path[fill=cebebeb,line cap=butt,line join=miter,line width=0.0231cm] (14.5786, 23.6324) -- (14.775, 23.4989) -- (14.7674, 22.9245) -- (14.5786, 22.8155) -- cycle;

  \path[fill=cebebeb,line cap=butt,line join=miter,line width=0.0231cm] (13.8237, 24.1441) -- (14.01, 24.0264) -- (14.0124, 23.3603) -- (13.8237, 23.2514) -- cycle;

  \path[fill=cebebeb,line cap=butt,line join=miter,line width=0.0231cm] (13.0687, 24.1488) -- (13.2651, 24.0307) -- (13.2575, 23.7962) -- (13.0687, 23.6873) -- cycle;

  \path[fill=cebebeb,line cap=butt,line join=miter,line width=0.0231cm] (12.3138, 24.876) -- (12.3138, 24.1231) -- (12.5025, 24.2321) -- (12.4963, 24.7484) -- cycle;

  \path[fill=cebebeb,line cap=butt,line join=miter,line width=0.0231cm] (11.5588, 24.559) -- (11.7476, 24.668) -- (11.7486, 24.7903) -- (11.5588, 24.9012) -- cycle;

  \path[fill=cebebeb,line cap=butt,line join=miter,line width=0.0231cm] (10.8039, 24.9949) -- (10.9926, 25.1038) -- (10.9856, 25.275) -- (10.8039, 25.3933) -- cycle;

  \path[fill=cebebeb,line cap=butt,line join=miter,line width=0.0231cm] (10.0489, 25.4307) -- (10.2377, 25.5397) -- (10.2344, 25.7332) -- (10.0489, 25.8292) -- cycle;

  \path[fill=cebebeb,line cap=butt,line join=miter,line width=0.0231cm] (9.294, 25.8666) -- (9.4827, 25.9756) -- (9.4737, 26.7051) -- (9.294, 26.8087) -- cycle;

  \path[fill=cebebeb,line cap=butt,line join=miter,line width=0.0231cm] (8.539, 26.3025) -- (8.7278, 26.4115) -- (8.7292, 27.033) -- (8.539, 27.1416) -- cycle;

  \path[fill=cebebeb,line cap=butt,line join=miter,line width=0.0231cm] (7.7841, 26.7384) -- (7.9728, 26.8473) -- (7.969, 27.0554) -- (7.7841, 27.1698) -- cycle;

  \path[fill=whitesmoke,line cap=butt,line join=miter,line width=0.0231cm] (15.3336, 22.3796) -- (15.5223, 22.4886) -- (16.0886, 22.1617) -- (16.0886, 21.9438) -- cycle;

  \path[fill=whitesmoke,line cap=butt,line join=miter,line width=0.0231cm] (14.5786, 22.8155) -- (14.7674, 22.9245) -- (15.3336, 22.5976) -- (15.3336, 22.3796) -- cycle;

  \path[fill=whitesmoke,line cap=butt,line join=miter,line width=0.0231cm] (13.8237, 23.2514) -- (14.0124, 23.3603) -- (14.5786, 23.0334) -- (14.5786, 22.8155) -- cycle;

  \path[fill=whitesmoke,line cap=butt,line join=miter,line width=0.0231cm] (13.0687, 23.6873) -- (13.2575, 23.7962) -- (13.8237, 23.4693) -- (13.8237, 23.2514) -- cycle;

  \path[fill=whitesmoke,line cap=butt,line join=miter,line width=0.0231cm] (12.3138, 24.1231) -- (12.5025, 24.2321) -- (13.0687, 23.9052) -- (13.0687, 23.6873) -- cycle;

  \path[fill=whitesmoke,line cap=butt,line join=miter,line width=0.0231cm] (11.5588, 24.559) -- (11.7476, 24.668) -- (12.3138, 24.3411) -- (12.3138, 24.1231) -- cycle;

  \path[fill=whitesmoke,line cap=butt,line join=miter,line width=0.0231cm] (10.8039, 24.9949) -- (10.9926, 25.1038) -- (11.5588, 24.7769) -- (11.5588, 24.559) -- cycle;

  \path[fill=whitesmoke,line cap=butt,line join=miter,line width=0.0231cm] (10.0489, 25.4307) -- (10.2377, 25.5397) -- (10.8039, 25.2128) -- (10.8039, 24.9949) -- cycle;

  \path[fill=whitesmoke,line cap=butt,line join=miter,line width=0.0231cm] (9.294, 25.8666) -- (9.4827, 25.9756) -- (10.0489, 25.6487) -- (10.0489, 25.4307) -- cycle;

  \path[fill=whitesmoke,line cap=butt,line join=miter,line width=0.0231cm] (8.539, 26.3025) -- (8.7278, 26.4115) -- (9.294, 26.0846) -- (9.294, 25.8666) -- cycle;

  \path[fill=whitesmoke,line cap=butt,line join=miter,line width=0.0231cm] (7.7841, 26.7384) -- (7.9728, 26.8473) -- (8.539, 26.5204) -- (8.539, 26.3025) -- cycle;

  \path[fill=gainsboro,line cap=butt,line join=miter,line width=0.0231cm] (7.7841, 26.7384) -- (16.0886, 21.9438) -- (16.0886, 21.2899) -- (7.7841, 26.0846) -- cycle;

  \path[fill=cebebeb,line cap=butt,line join=miter,line width=0.0231cm] (4.7642, 26.3025) -- (7.7841, 28.046) -- (7.7841, 26.0846) -- (4.7642, 24.3411) -- cycle;

  \path[draw=c191966,fill=whitesmoke,line cap=butt,line join=miter,line width=0.0231cm,miter limit=4.0] (4.7642, 24.3411) -- (13.0687, 19.5465) -- (16.0886, 21.2899) -- (7.7841, 26.0846) -- (4.7642, 24.3411);

  \path[draw=black,even odd rule,line cap=butt,line join=miter,line width=0.0231cm] (4.7642, 24.3411) -- (13.0687, 19.5465) -- (16.2773, 21.3989) -- (16.2773, 23.3603) -- (7.9728, 28.1549) -- (4.7642, 26.3025) -- cycle(8.1615, 27.828) -- (15.7111, 23.4693);

  \path[draw=black,even odd rule,line cap=butt,line join=miter,line width=0.0231cm,miter limit=4.0] (4.7642, 24.3411) -- (7.7841, 26.0846) -- (7.7841, 28.046);

  \path[draw=black,even odd rule,line cap=butt,line join=miter,line width=0.0154cm,miter limit=4.0] (7.7841, 26.0846) -- (16.0886, 21.2899);

  \path[draw=black,even odd rule,line cap=butt,line join=miter,line width=0.0231cm] (7.7841, 26.7384) -- (16.0886, 21.9438);

  \path[draw=black,even odd rule,line cap=butt,line join=miter,line width=0.0231cm] (8.539, 27.6101) -- (8.539, 26.3025);

  \path[draw=black,even odd rule,line cap=butt,line join=miter,line width=0.0231cm] (9.294, 27.1742) -- (9.294, 25.8666);

  \path[draw=black,even odd rule,line cap=butt,line join=miter,line width=0.0231cm] (10.0489, 26.7384) -- (10.0489, 25.4307);

  \path[draw=black,even odd rule,line cap=butt,line join=miter,line width=0.0231cm] (10.8039, 26.3025) -- (10.8039, 24.9949);

  \path[draw=black,even odd rule,line cap=butt,line join=miter,line width=0.0231cm] (11.5588, 25.8666) -- (11.5588, 24.559);

  \path[draw=black,even odd rule,line cap=butt,line join=miter,line width=0.0231cm] (12.3138, 25.4307) -- (12.3138, 24.1231);

  \path[draw=black,even odd rule,line cap=butt,line join=miter,line width=0.0231cm] (13.0687, 24.9949) -- (13.0687, 23.6873);

  \path[draw=black,even odd rule,line cap=butt,line join=miter,line width=0.0231cm] (13.8237, 24.559) -- (13.8237, 23.2514);

  \path[draw=black,even odd rule,line cap=butt,line join=miter,line width=0.0231cm] (14.5786, 24.1231) -- (14.5786, 22.8155);

  \path[draw=black,even odd rule,line cap=butt,line join=miter,line width=0.0231cm] (15.3336, 23.6873) -- (15.3336, 22.3796);

  \path[draw=black,even odd rule,line cap=butt,line join=miter,line width=0.0154cm,miter limit=4.0] (7.9728, 28.1549) -- (7.9728, 26.8473) -- (7.7841, 26.7384);

  \path[draw=black,even odd rule,line cap=butt,line join=miter,line width=0.0154cm,miter limit=4.0] (7.9728, 26.8473) -- (8.539, 26.5204);

  \path[draw=black,even odd rule,line cap=butt,line join=miter,line width=0.0231cm] (8.539, 27.6101) -- (8.7278, 27.7191);

  \path[draw=black,even odd rule,line cap=butt,line join=miter,line width=0.0231cm] (9.294, 27.1742) -- (9.4827, 27.2832);

  \path[draw=black,even odd rule,line cap=butt,line join=miter,line width=0.0231cm] (10.0489, 26.7384) -- (10.2377, 26.8473);

  \path[draw=black,even odd rule,line cap=butt,line join=miter,line width=0.0154cm,miter limit=4.0] (8.7278, 27.7191) -- (8.7278, 26.4115) -- (8.539, 26.3025);

  \path[draw=black,even odd rule,line cap=butt,line join=miter,line width=0.0154cm,miter limit=4.0] (8.7278, 26.4115) -- (9.294, 26.0846);

  \path[draw=black,even odd rule,line cap=butt,line join=miter,line width=0.0153cm,miter limit=4.0] (9.4827, 27.2832) -- (9.4827, 25.9756) -- (9.294, 25.8666);

  \path[draw=black,even odd rule,line cap=butt,line join=miter,line width=0.0231cm] (9.4827, 25.9756) -- (10.0489, 25.6487);

  \path[draw=black,even odd rule,line cap=butt,line join=miter,line width=0.0154cm,miter limit=4.0] (10.2377, 26.8473) -- (10.2377, 25.5397) -- (10.0489, 25.4307);

  \path[draw=black,even odd rule,line cap=butt,line join=miter,line width=0.0154cm,miter limit=4.0] (10.2377, 25.5397) -- (10.8039, 25.2128);

  \path[draw=black,even odd rule,line cap=butt,line join=miter,line width=0.0154cm,miter limit=4.0] (10.9926, 26.4115) -- (10.9926, 25.1038) -- (10.8039, 24.9949);

  \path[draw=black,even odd rule,line cap=butt,line join=miter,line width=0.0154cm,miter limit=4.0] (10.9926, 25.1038) -- (11.5588, 24.7769);

  \path[draw=black,even odd rule,line cap=butt,line join=miter,line width=0.0154cm,miter limit=4.0] (11.7476, 25.9756) -- (11.7476, 24.668) -- (11.5588, 24.559);

  \path[draw=black,even odd rule,line cap=butt,line join=miter,line width=0.0154cm,miter limit=4.0] (11.7476, 24.668) -- (12.3138, 24.3411);

  \path[draw=black,even odd rule,line cap=butt,line join=miter,line width=0.0154cm,miter limit=4.0] (12.5025, 25.5397) -- (12.5025, 24.2321) -- (12.3138, 24.1231);

  \path[draw=black,even odd rule,line cap=butt,line join=miter,line width=0.0154cm,miter limit=4.0] (12.5025, 24.2321) -- (13.0687, 23.9052);

  \path[draw=black,even odd rule,line cap=butt,line join=miter,line width=0.0154cm,miter limit=4.0] (13.2575, 25.1038) -- (13.2575, 23.7962) -- (13.0687, 23.6873);

  \path[draw=black,even odd rule,line cap=butt,line join=miter,line width=0.0154cm,miter limit=4.0] (13.2575, 23.7962) -- (13.8237, 23.4693);

  \path[draw=black,even odd rule,line cap=butt,line join=miter,line width=0.0154cm,miter limit=4.0] (14.0124, 24.668) -- (14.0124, 23.3603) -- (13.8237, 23.2514);

  \path[draw=black,even odd rule,line cap=butt,line join=miter,line width=0.0154cm,miter limit=4.0] (14.0124, 23.3603) -- (14.5786, 23.0334);

  \path[draw=black,even odd rule,line cap=butt,line join=miter,line width=0.0154cm,miter limit=4.0] (14.7674, 24.2321) -- (14.7674, 22.9245) -- (14.5786, 22.8155);

  \path[draw=black,even odd rule,line cap=butt,line join=miter,line width=0.0154cm,miter limit=4.0] (14.7674, 22.9245) -- (15.3336, 22.5976);

  \path[draw=black,even odd rule,line cap=butt,line join=miter,line width=0.0154cm,miter limit=4.0] (15.5223, 23.7962) -- (15.5223, 22.4886) -- (15.3336, 22.3796);

  \path[draw=black,even odd rule,line cap=butt,line join=miter,line width=0.0154cm,miter limit=4.0] (15.5223, 22.4886) -- (16.0886, 22.1617);

  \path[draw=black,even odd rule,line cap=butt,line join=miter,line width=0.0231cm] (10.8039, 26.3025) -- (10.9926, 26.4115);

  \path[draw=black,even odd rule,line cap=butt,line join=miter,line width=0.0231cm] (11.5588, 25.8666) -- (11.7476, 25.9756);

  \path[draw=black,even odd rule,line cap=butt,line join=miter,line width=0.0231cm] (12.3138, 25.4307) -- (12.5025, 25.5397);

  \path[draw=black,even odd rule,line cap=butt,line join=miter,line width=0.0231cm] (13.0687, 24.9949) -- (13.2575, 25.1038);

  \path[draw=black,even odd rule,line cap=butt,line join=miter,line width=0.0231cm] (13.8237, 24.559) -- (14.0124, 24.668);

  \path[draw=black,even odd rule,line cap=butt,line join=miter,line width=0.0231cm] (14.5786, 24.1231) -- (14.7674, 24.2321);

  \path[draw=black,even odd rule,line cap=butt,line join=miter,line width=0.0231cm] (15.3336, 23.6873) -- (15.5223, 23.7962);

  \path[draw=black,even odd rule,line cap=butt,line join=miter,line width=0.0231cm] (7.7841, 28.046) -- (8.539, 27.6101) -- (8.539, 26.7339) -- (7.7841, 27.1698) -- (7.7841, 28.046);

  \path[draw=black,even odd rule,line cap=butt,line join=miter,line width=0.0231cm] (8.539, 27.6101) -- (9.294, 27.1742) -- (9.294, 26.7058) -- (8.539, 27.1416) -- (8.539, 27.6101);

  \path[draw=black,even odd rule,line cap=butt,line join=miter,line width=0.0231cm] (9.294, 27.1742) -- (10.0489, 26.7384) -- (10.0489, 26.3729) -- (9.294, 26.8087) -- (9.294, 27.1742);

  \path[draw=black,even odd rule,line cap=butt,line join=miter,line width=0.0231cm] (10.0489, 26.7384) -- (10.8039, 26.3025) -- (10.8039, 25.3933) -- (10.0489, 25.8292) -- (10.0489, 26.7384);

  \path[draw=black,even odd rule,line cap=butt,line join=miter,line width=0.0231cm] (10.8039, 26.3025) -- (11.5588, 25.8666) -- (11.5588, 24.9536) -- (10.8039, 25.3895) -- (10.8039, 26.3025);

  \path[draw=black,even odd rule,line cap=butt,line join=miter,line width=0.0231cm] (11.5588, 25.8666) -- (12.3138, 25.4307) -- (12.3138, 24.4653) -- (11.5588, 24.9012) -- (11.5588, 25.8666);

  \path[draw=black,even odd rule,line cap=butt,line join=miter,line width=0.0231cm] (12.3138, 25.4307) -- (13.0687, 24.9949) -- (13.0687, 24.4401) -- (12.3138, 24.876) -- (12.3138, 25.4307);

  \path[draw=black,even odd rule,line cap=butt,line join=miter,line width=0.0231cm] (13.0687, 24.9949) -- (13.8237, 24.559) -- (13.8237, 23.713) -- (13.0687, 24.1488) -- (13.0687, 24.9949);

  \path[draw=black,even odd rule,line cap=butt,line join=miter,line width=0.0231cm] (13.8237, 24.559) -- (14.5786, 24.1231) -- (14.5786, 23.7082) -- (13.8237, 24.1441) -- (13.8237, 24.559);

  \path[draw=black,even odd rule,line cap=butt,line join=miter,line width=0.0231cm] (14.5786, 24.1231) -- (15.3336, 23.6873) -- (15.3336, 23.1965) -- (14.5786, 23.6324) -- (14.5786, 24.1231);

  \path[draw=black,even odd rule,line cap=butt,line join=miter,line width=0.0231cm] (15.3336, 23.6873) -- (16.0886, 23.2514) -- (16.0886, 22.8859) -- (15.3336, 23.3218) -- (15.3336, 23.6873);

  \path[draw=black,even odd rule,line cap=butt,line join=miter,line width=0.0231cm] (16.0886, 21.9438) -- (16.0886, 23.2514);

  \path[draw=black,fill=white,even odd rule,line cap=butt,line join=miter,line width=0.0231cm] (9.1052, 25.5397) -- (7.4066, 24.559) -- (7.9728, 24.2321) -- (9.6714, 25.2128) -- cycle;

  \path[draw=black,fill=white,even odd rule,line cap=butt,line join=miter,line width=0.0231cm] (10.2377, 24.8859) -- (8.539, 23.9052) -- (9.1052, 23.5783) -- (10.8039, 24.559) -- cycle;

  \path[draw=black,fill=white,even odd rule,line cap=butt,line join=miter,line width=0.0231cm] (11.3701, 24.2321) -- (9.6714, 23.2514) -- (10.2377, 22.9245) -- (11.9363, 23.9052) -- cycle;

  \path[draw=black,fill=white,even odd rule,line cap=butt,line join=miter,line width=0.0231cm] (12.5025, 23.5783) -- (10.8039, 22.5976) -- (11.3701, 22.2707) -- (13.0687, 23.2514) -- cycle;

  \path[draw=black,fill=white,even odd rule,line cap=butt,line join=miter,line width=0.0231cm] (13.635, 22.9245) -- (11.9363, 21.9438) -- (12.5025, 21.6169) -- (14.2012, 22.5976) -- cycle;

  \path[draw=black,fill=white,even odd rule,line cap=butt,line join=miter,line width=0.0231cm] (14.7674, 22.2707) -- (13.0687, 21.2899) -- (13.635, 20.963) -- (15.3336, 21.9438) -- cycle;

  \path[draw=black,fill=white,even odd rule,line cap=butt,line join=miter,line width=0.0231cm] (9.1052, 25.5397) -- (7.2178, 24.45) -- (7.7841, 24.1231) -- (9.6714, 25.2128) -- cycle;

  \path[draw=black,even odd rule,line cap=butt,line join=miter,line width=0.0231cm] (9.1052, 25.5397) -- (9.6714, 25.2128);

  \path[draw=black,even odd rule,line cap=butt,line join=miter,line width=0.0154cm,miter limit=4.0] (7.501, 24.2866) -- (9.3883, 25.3763);

  \path[draw=black,even odd rule,line cap=butt,line join=miter,line width=0.0154cm,miter limit=4.0] (8.7278, 25.3218) -- (9.294, 24.9949);

  \path[draw=black,even odd rule,line cap=butt,line join=miter,line width=0.0154cm,miter limit=4.0] (8.3503, 25.1038) -- (8.9165, 24.7769);

  \path[draw=black,even odd rule,line cap=butt,line join=miter,line width=0.0154cm,miter limit=4.0] (7.9728, 24.8859) -- (8.539, 24.559);

  \path[draw=black,even odd rule,line cap=butt,line join=miter,line width=0.0154cm,miter limit=4.0] (7.5953, 24.668) -- (8.1615, 24.3411);

  \path[draw=black,fill=white,even odd rule,line cap=butt,line join=miter,line width=0.0231cm] (10.2377, 24.8859) -- (8.3503, 23.7962) -- (8.9165, 23.4693) -- (10.8039, 24.559) -- cycle;

  \path[draw=black,even odd rule,line cap=butt,line join=miter,line width=0.0231cm] (10.2377, 24.8859) -- (10.8039, 24.559);

  \path[draw=black,even odd rule,line cap=butt,line join=miter,line width=0.0154cm,miter limit=4.0] (8.6334, 23.6328) -- (10.5208, 24.7224);

  \path[draw=black,even odd rule,line cap=butt,line join=miter,line width=0.0154cm,miter limit=4.0] (9.8602, 24.668) -- (10.4264, 24.3411);

  \path[draw=black,even odd rule,line cap=butt,line join=miter,line width=0.0154cm,miter limit=4.0] (9.4827, 24.45) -- (10.0489, 24.1231);

  \path[draw=black,even odd rule,line cap=butt,line join=miter,line width=0.0154cm,miter limit=4.0] (9.1052, 24.2321) -- (9.6714, 23.9052);

  \path[draw=black,even odd rule,line cap=butt,line join=miter,line width=0.0154cm,miter limit=4.0] (8.7278, 24.0142) -- (9.294, 23.6873);

  \path[draw=black,fill=white,even odd rule,line cap=butt,line join=miter,line width=0.0231cm] (12.5025, 23.5783) -- (10.6151, 22.4886) -- (11.1814, 22.1617) -- (13.0687, 23.2514) -- cycle;

  \path[draw=black,even odd rule,line cap=butt,line join=miter,line width=0.0231cm] (12.5025, 23.5783) -- (13.0687, 23.2514);

  \path[draw=black,even odd rule,line cap=butt,line join=miter,line width=0.0154cm,miter limit=4.0] (10.8982, 22.3251) -- (12.7856, 23.4148);

  \path[draw=black,even odd rule,line cap=butt,line join=miter,line width=0.0154cm,miter limit=4.0] (12.125, 23.3603) -- (12.6913, 23.0334);

  \path[draw=black,even odd rule,line cap=butt,line join=miter,line width=0.0154cm,miter limit=4.0] (11.7476, 23.1424) -- (12.3138, 22.8155);

  \path[draw=black,even odd rule,line cap=butt,line join=miter,line width=0.0154cm,miter limit=4.0] (11.3701, 22.9245) -- (11.9363, 22.5976);

  \path[draw=black,even odd rule,line cap=butt,line join=miter,line width=0.0154cm,miter limit=4.0] (10.9926, 22.7065) -- (11.5588, 22.3796);

  \path[draw=black,fill=white,even odd rule,line cap=butt,line join=miter,line width=0.0231cm] (13.635, 22.9245) -- (11.7476, 21.8348) -- (12.3138, 21.5079) -- (14.2012, 22.5976) -- cycle;

  \path[draw=black,even odd rule,line cap=butt,line join=miter,line width=0.0231cm] (13.635, 22.9245) -- (14.2012, 22.5976);

  \path[draw=black,even odd rule,line cap=butt,line join=miter,line width=0.0154cm,miter limit=4.0] (12.0307, 21.6713) -- (13.9181, 22.761);

  \path[draw=black,even odd rule,line cap=butt,line join=miter,line width=0.0154cm,miter limit=4.0] (13.2575, 22.7065) -- (13.8237, 22.3796);

  \path[draw=black,even odd rule,line cap=butt,line join=miter,line width=0.0154cm,miter limit=4.0] (12.88, 22.4886) -- (13.4462, 22.1617);

  \path[draw=black,even odd rule,line cap=butt,line join=miter,line width=0.0154cm,miter limit=4.0] (12.5025, 22.2707) -- (13.0687, 21.9438);

  \path[draw=black,even odd rule,line cap=butt,line join=miter,line width=0.0154cm,miter limit=4.0] (12.125, 22.0527) -- (12.6913, 21.7258);

  \path[draw=black,even odd rule,line cap=butt,line join=miter,line width=0.0231cm] (6.0854, 25.1038) -- (6.6516, 24.7769);

  \path[draw=black,fill=white,even odd rule,line cap=butt,line join=miter,line width=0.0231cm] (7.9728, 26.1935) -- (6.0854, 25.1038) -- (6.6516, 24.7769) -- (8.539, 25.8666) -- cycle;

  \path[draw=black,even odd rule,line cap=butt,line join=miter,line width=0.0231cm] (7.9728, 26.1935) -- (8.539, 25.8666);

  \path[draw=black,even odd rule,line cap=butt,line join=miter,line width=0.0154cm,miter limit=4.0] (6.3685, 24.9404) -- (8.2559, 26.0301);

  \path[draw=black,even odd rule,line cap=butt,line join=miter,line width=0.0154cm,miter limit=4.0] (7.5953, 25.9756) -- (8.1615, 25.6487);

  \path[draw=black,even odd rule,line cap=butt,line join=miter,line width=0.0154cm,miter limit=4.0] (7.2178, 25.7576) -- (7.7841, 25.4307);

  \path[draw=black,even odd rule,line cap=butt,line join=miter,line width=0.0154cm,miter limit=4.0] (6.8404, 25.5397) -- (7.4066, 25.2128);

  \path[draw=black,even odd rule,line cap=butt,line join=miter,line width=0.0154cm,miter limit=4.0] (6.4629, 25.3218) -- (7.0291, 24.9949);

  \path[draw=black,fill=white,even odd rule,line cap=butt,line join=miter,line width=0.0231cm] (11.3701, 24.2321) -- (9.4827, 23.1424) -- (10.0489, 22.8155) -- (11.9363, 23.9052) -- cycle;

  \path[draw=black,even odd rule,line cap=butt,line join=miter,line width=0.0231cm] (11.3701, 24.2321) -- (11.9363, 23.9052);

  \path[draw=black,even odd rule,line cap=butt,line join=miter,line width=0.0154cm,miter limit=4.0] (9.7658, 22.979) -- (11.6532, 24.0686);

  \path[draw=black,even odd rule,line cap=butt,line join=miter,line width=0.0154cm,miter limit=4.0] (10.9926, 24.0142) -- (11.5588, 23.6873);

  \path[draw=black,even odd rule,line cap=butt,line join=miter,line width=0.0154cm,miter limit=4.0] (10.6151, 23.7962) -- (11.1814, 23.4693);

  \path[draw=black,even odd rule,line cap=butt,line join=miter,line width=0.0154cm,miter limit=4.0] (10.2377, 23.5783) -- (10.8039, 23.2514);

  \path[draw=black,even odd rule,line cap=butt,line join=miter,line width=0.0154cm,miter limit=4.0] (9.8602, 23.3603) -- (10.4264, 23.0334);

  \path[draw=black,fill=white,even odd rule,line cap=butt,line join=miter,line width=0.0231cm] (14.7674, 22.2707) -- (12.88, 21.181) -- (13.4462, 20.8541) -- (15.3336, 21.9438) -- cycle;

  \path[draw=black,even odd rule,line cap=butt,line join=miter,line width=0.0231cm] (14.7674, 22.2707) -- (15.3336, 21.9438);

  \path[draw=black,even odd rule,line cap=butt,line join=miter,line width=0.0154cm,miter limit=4.0] (13.1631, 21.0175) -- (15.0505, 22.1072);

  \path[draw=black,even odd rule,line cap=butt,line join=miter,line width=0.0154cm,miter limit=4.0] (14.3899, 22.0527) -- (14.9561, 21.7258);

  \path[draw=black,even odd rule,line cap=butt,line join=miter,line width=0.0154cm,miter limit=4.0] (14.0124, 21.8348) -- (14.5786, 21.5079);

  \path[draw=black,even odd rule,line cap=butt,line join=miter,line width=0.0154cm,miter limit=4.0] (13.635, 21.6169) -- (14.2012, 21.2899);

  \path[draw=black,even odd rule,line cap=butt,line join=miter,line width=0.0154cm,miter limit=4.0] (13.2575, 21.3989) -- (13.8237, 21.072);

  \path[draw=black,fill=white,even odd rule,line cap=butt,line join=miter,line width=0.0231cm] (4.953, 24.2321) -- (4.953, 25.3218) -- (5.3305, 25.1038) -- (5.3305, 24.0142);

  \path[draw=black,fill=white,even odd rule,line cap=butt,line join=miter,line width=0.0231cm] (12.5025, 19.8734) -- (12.5025, 20.963) -- (12.88, 20.7451) -- (12.88, 19.6554);

  \path[draw=black,even odd rule,line cap=butt,line join=miter,line width=0.0123cm,miter limit=4.0] (8.9165, 24.1231) -- (8.539, 23.9052);

  \path[draw=black,even odd rule,line cap=butt,line join=miter,line width=0.0123cm,miter limit=4.0] (8.9165, 24.7769) -- (8.539, 24.559);

  \path[fill=cebebeb,line cap=butt,line join=miter,line width=0.0231cm] (11.9363, 20.6328) -- (12.3138, 20.8507) -- (12.3166, 22.6095) -- (12.1378, 22.7145) -- (11.9363, 22.5942) -- cycle;

  \path[fill=white,fill opacity=0.9957,line cap=butt,line join=miter,line width=0.0231cm] (11.5588, 20.8507) -- (11.9363, 20.6328) -- (11.9363, 22.5976) -- (11.5588, 22.8155) -- cycle;

  \path[draw=black,fill opacity=0.9957,even odd rule,line cap=butt,line join=miter,line width=0.0231cm] (11.9363, 20.6328) -- (11.9363, 22.5942) -- (11.5588, 22.8122) -- (11.5588, 20.8507) -- (11.9363, 20.6328) -- (12.3138, 20.8507) -- (12.3138, 22.8122) -- (11.9363, 22.5942);

  \path[draw=black,even odd rule,line cap=butt,line join=miter,line width=0.0123cm,miter limit=4.0] (5.5192, 25.8666) -- (5.7079, 25.7576);

  \path[draw=black,even odd rule,line cap=butt,line join=miter,line width=0.0123cm,miter limit=4.0] (10.2377, 24.2321) -- (9.8602, 24.0142);

  \path[draw=black,fill=white,fill opacity=0.9957,even odd rule,line cap=butt,line join=miter,line width=0.0231cm] (11.5588, 22.8155) -- (11.9363, 22.5976) -- (12.3138, 22.8155) -- (11.9363, 23.0334) -- cycle;

  \path[draw=black,even odd rule,line cap=butt,line join=miter,line width=0.0231cm,miter limit=4.0,dash pattern=on 0.1848cm off 0.0924cm] (4.7642, 26.3025) -- (13.0687, 21.5079) -- (16.2773, 23.3603);

  \path[draw=black,even odd rule,line cap=butt,line join=miter,line width=0.0231cm] (16.0886, 21.9438) -- (16.0886, 21.2899);

  \path[draw=black,even odd rule,line cap=butt,line join=miter,line width=0.0231cm,miter limit=4.0,dash pattern=on 0.1845cm off 0.1845cm] (13.0687, 21.5079) -- (13.0687, 19.5465);

  \path[fill=white,fill opacity=0.4915,line cap=round,line join=round,line width=0.0185cm] (4.2101, 28.3791) rectangle (17.2075, 19.119);

  \begin{scope}[shift={(-0.0149, 0.2144)}]
    \path[draw=black,fill=cecdf00,line cap=round,line join=round,line width=0.0106cm] (12.3817, 22.425) circle (0.0869cm);

    \path[draw=black,fill=cecdf00,line cap=round,line join=round,line width=0.0106cm] (5.7447, 26.4969) circle (0.0869cm);

    \path[draw=black,fill=cecdf00,line cap=round,line join=round,line width=0.0106cm] (5.7463, 26.0001) circle (0.0869cm);

    \path[draw=black,fill=cecdf00,line cap=round,line join=round,line width=0.0106cm] (12.358, 21.8597) circle (0.0869cm);

    \path[draw=black,fill=cecdf00,line cap=round,line join=round,line width=0.0106cm] (10.8228, 25.6468) circle (0.0869cm);

    \path[draw=black,fill=cecdf00,line cap=round,line join=round,line width=0.0106cm] (12.3482, 21.2117) circle (0.0869cm);

    \path[draw=black,fill=cecdf00,line cap=round,line join=round,line width=0.0106cm] (12.3557, 20.8925) circle (0.0869cm);

    \node[text=black,anchor=south west,line cap=round,line join=round,line width=0.0141cm] (text1) at (5.9352, 26.4373){$S_{11}$};

    \node[text=black,anchor=south west,line cap=round,line join=round,line width=0.0141cm] (text1-2) at (5.9275, 25.9627){$S_{12}$};

    \node[text=black,anchor=south west,line cap=round,line join=round,line width=0.0141cm] (text1-2-5) at (11.8525, 26.1684){$S_{22}$};

    \node[text=black,anchor=south west,line cap=round,line join=round,line width=0.0141cm] (text1-2-5-9) at (10.9172, 21.6937){$S_{31}$};

    \node[text=black,anchor=south west,line cap=round,line join=round,line width=0.0141cm] (text1-2-5-9-2) at (10.8843, 21.2218){$S_{32}$};

    \node[text=black,anchor=south west,line cap=round,line join=round,line width=0.0141cm] (text1-2-5-9-2-4) at (10.9003, 20.624){$S_{33}$};

    \node[text=black,anchor=south west,line cap=round,line join=round,line width=0.0141cm] (text1-2-5-9-2-4-0) at (10.895, 20.2405){$S_{34}$};

    \path[draw=black,line cap=round,line join=round,line width=0.0141cm,dash pattern=on 0.0564cm off 0.0564cm] (11.4531, 20.3607) -- (12.2429, 20.8453);

    \path[draw=black,line cap=round,line join=round,line width=0.0141cm,dash pattern=on 0.0564cm off 0.0564cm] (11.4401, 20.6937) -- (12.2537, 21.1635);

    \path[draw=black,line cap=round,line join=round,line width=0.0141cm,dash pattern=on 0.0564cm off 0.0564cm] (11.4367, 21.3392) -- (12.2503, 21.8089);

    \path[draw=black,line cap=round,line join=round,line width=0.0141cm,dash pattern=on 0.0564cm off 0.0564cm] (11.4652, 21.9255) -- (12.2788, 22.3952);

    \path[draw=black,line cap=round,line join=round,line width=0.0141cm,dash pattern=on 0.0564cm off 0.0564cm] (10.93, 25.704) -- (11.7436, 26.1737);

  \end{scope}
  \path[draw=black,fill=cd24444,even odd rule,line cap=butt,line join=miter,line width=0.0231cm,miter limit=4.0] (7.5318, 26.632) -- (7.9093, 26.85) -- (8.2868, 26.632) -- (7.9093, 26.4141) -- cycle;

  \path[draw=c3a46a3,fill=white,fill opacity=0.5032,line cap=round,line join=round,line width=0.0923cm] (7.3834, 26.2712) -- (8.9809, 27.2152);

  \path[draw=c3a46a3,fill=white,fill opacity=0.5032,line cap=round,line join=round,line width=0.0923cm] (7.8437, 26.0378) -- (9.4412, 26.9818);

  \path[draw=black,fill=cd24444,even odd rule,line cap=butt,line join=miter,line width=0.0231cm,miter limit=4.0] (9.2892, 25.6343) -- (9.6666, 25.8522) -- (10.0441, 25.6343) -- (9.6666, 25.4164) -- cycle;

  \path[draw=c3a46a3,fill=white,fill opacity=0.5032,line cap=round,line join=round,line width=0.0923cm] (9.1345, 25.2257) -- (10.732, 26.1697);

  \path[draw=c3a46a3,fill=white,fill opacity=0.5032,line cap=round,line join=round,line width=0.0923cm] (9.5948, 24.9923) -- (11.1923, 25.9363);

  \path[draw=black,fill=cd24444,even odd rule,line cap=butt,line join=miter,line width=0.0231cm,miter limit=4.0] (10.8175, 24.7369) -- (11.1949, 24.9548) -- (11.5724, 24.7369) -- (11.1949, 24.5189) -- cycle;

  \path[draw=c3a46a3,fill=white,fill opacity=0.5032,line cap=round,line join=round,line width=0.0923cm] (10.6856, 24.334) -- (12.2831, 25.278);

  \path[draw=black,fill=cd24444,even odd rule,line cap=butt,line join=miter,line width=0.0231cm,miter limit=4.0] (12.397, 23.8089) -- (12.7744, 24.0268) -- (13.1519, 23.8089) -- (12.7744, 23.591) -- cycle;

  \path[draw=c3a46a3,fill=white,fill opacity=0.5032,line cap=round,line join=round,line width=0.0923cm] (11.146, 24.1007) -- (12.7435, 25.0447);

  \path[draw=c3a46a3,fill=white,fill opacity=0.5032,line cap=round,line join=round,line width=0.0923cm] (12.5187, 23.3117) -- (14.1162, 24.2557);

  \path[draw=c3a46a3,fill=white,fill opacity=0.5032,line cap=round,line join=round,line width=0.0923cm] (8.3274, 27.6343) -- (14.3556, 24.141);

  \path[draw=c3a46a3,fill=white,fill opacity=0.5032,line cap=round,line join=round,line width=0.0923cm] (5.5106, 26.0786) -- (11.7703, 22.4885);

  \path[draw=black,fill=c9da5dd,even odd rule,line cap=butt,line join=miter,line width=0.0231cm] (7.41, 25.4481) -- (7.9763, 25.775) -- (7.9763, 25.993) -- (7.2213, 26.4288) -- (6.6551, 26.1019) -- (6.6551, 25.884) -- cycle;

  \path[draw=black,even odd rule,line cap=butt,line join=miter,line width=0.0154cm,miter limit=4.0] (7.41, 25.4481) -- (7.41, 25.6661) -- (7.9763, 25.993);

  \path[draw=c3a46a3,fill opacity=0.5032,line cap=round,line join=round,line width=0.0923cm] (7.1813, 25.1602) -- (7.7361, 25.4744) -- (7.5335, 25.6083);

  \path[draw=black,fill=cd24444,even odd rule,line cap=butt,line join=miter,line width=0.0231cm,miter limit=4.0] (14.026, 22.8844) -- (14.4035, 23.1023) -- (14.781, 22.8844) -- (14.4035, 22.6665) -- cycle;

  \path[draw=black,even odd rule,line cap=butt,line join=miter,line width=0.0154cm,miter limit=4.0] (7.41, 25.6661) -- (6.6551, 26.1019);

  \path[draw=black,fill=c9da5dd,even odd rule,line cap=butt,line join=miter,line width=0.0231cm] (9.2049, 24.4031) -- (9.7711, 24.73) -- (9.7711, 24.9479) -- (9.0161, 25.3838) -- (8.4499, 25.0569) -- (8.4499, 24.8389) -- cycle;

  \path[draw=black,even odd rule,line cap=butt,line join=miter,line width=0.0154cm,miter limit=4.0] (9.2049, 24.4031) -- (9.2049, 24.621) -- (9.7711, 24.9479);

  \path[draw=c3a46a3,fill opacity=0.5032,line cap=round,line join=round,line width=0.0923cm] (9.0363, 24.0987) -- (9.5911, 24.4129) -- (9.3885, 24.5468);

  \path[draw=black,even odd rule,line cap=butt,line join=miter,line width=0.0154cm,miter limit=4.0] (9.2049, 24.621) -- (8.4499, 25.0569);

  \path[draw=black,fill=c9da5dd,even odd rule,line cap=butt,line join=miter,line width=0.0231cm] (10.7387, 23.5331) -- (11.3049, 23.86) -- (11.3049, 24.0779) -- (10.5499, 24.5138) -- (9.9837, 24.1869) -- (9.9837, 23.969) -- cycle;

  \path[draw=c3a46a3,fill opacity=0.5032,line cap=round,line join=round,line width=0.0923cm] (10.563, 23.2372) -- (11.1178, 23.5514) -- (10.9151, 23.6853);

  \path[draw=black,even odd rule,line cap=butt,line join=miter,line width=0.0154cm,miter limit=4.0] (10.7387, 23.5331) -- (10.7387, 23.751) -- (11.3049, 24.0779);

  \path[draw=black,even odd rule,line cap=butt,line join=miter,line width=0.0154cm,miter limit=4.0] (10.7387, 23.751) -- (9.9837, 24.1869);

  \path[draw=ca82929,fill=cd24444,line cap=round,line join=round,line width=0.1845cm] (7.9122, 26.7323) -- (7.9122, 27.1926);

  \path[draw=ca82929,fill=cd24444,line cap=round,line join=round,line width=0.1845cm] (9.6929, 25.7081) -- (9.6929, 26.1684);

  \path[draw=ca82929,fill=cd24444,line cap=round,line join=round,line width=0.1845cm] (11.2326, 24.8398) -- (11.2326, 25.3002);

  \path[draw=ca82929,fill=cd24444,line cap=round,line join=round,line width=0.1845cm] (12.7805, 23.8959) -- (12.7805, 24.3563);

  \path[draw=ca82929,fill=cd24444,line cap=round,line join=round,line width=0.1845cm] (14.4317, 22.9767) -- (14.4317, 23.4371);

  \path[draw=c3a46a3,fill=white,fill opacity=0.5032,line cap=round,line join=round,line width=0.0923cm] (12.1089, 23.524) -- (13.7064, 24.4681);

  \path[draw=black,fill=c9da5dd,even odd rule,line cap=butt,line join=miter,line width=0.0231cm] (12.1201, 22.7473) -- (12.6864, 23.0742) -- (12.6864, 23.2922) -- (11.9314, 23.728) -- (11.3652, 23.4011) -- (11.3652, 23.1832) -- cycle;

  \path[draw=ca82929,line cap=round,line join=round,line width=0.1845cm] (7.4501, 27.958) -- (7.4487, 27.4836) -- (14.4302, 23.4713);

  \path[draw=ca82929,fill opacity=0.5242,line cap=round,line join=round,line width=0.0308cm] (7.4509, 28.1169) -- (7.4509, 28.469);

  \path[draw=ca82929,fill opacity=0.5242,line cap=round,line join=round,line width=0.0308cm] (7.3803, 28.3563) -- (7.4499, 28.4714) -- (7.5181, 28.3603);

  \path[draw=cc2c2db,fill opacity=0.5242,line cap=round,line join=round,line width=0.0308cm] (6.0476, 28.4225) -- (6.0476, 28.0705);

  \path[draw=cc2c2db,fill opacity=0.5242,line cap=round,line join=round,line width=0.0308cm] (5.977, 28.1832) -- (6.0466, 28.0681) -- (6.1148, 28.1791);

  \path[draw=c3a46a3,fill opacity=0.5242,line cap=round,line join=round,line width=0.0185cm] (5.9093, 25.7429) -- (5.9093, 25.4936);

  \path[draw=c3a46a3,fill opacity=0.5242,line cap=round,line join=round,line width=0.0185cm] (5.8593, 25.5734) -- (5.9086, 25.4919) -- (5.9569, 25.5705);

  \path[draw=c3a46a3,fill opacity=0.5242,line cap=round,line join=round,line width=0.0185cm] (6.4022, 25.4445) -- (6.4022, 25.1952);

  \path[draw=c3a46a3,fill opacity=0.5242,line cap=round,line join=round,line width=0.0185cm] (6.3522, 25.275) -- (6.4015, 25.1935) -- (6.4498, 25.2722);

  \path[draw=c3a46a3,fill opacity=0.5242,line cap=round,line join=round,line width=0.0185cm] (6.865, 25.1933) -- (6.865, 24.944);

  \path[draw=c3a46a3,fill opacity=0.5242,line cap=round,line join=round,line width=0.0185cm] (6.815, 25.0238) -- (6.8643, 24.9423) -- (6.9126, 25.0209);

  \path[draw=c3a46a3,fill opacity=0.5242,line cap=round,line join=round,line width=0.0185cm] (7.3821, 24.9103) -- (7.3821, 24.661);

  \path[draw=c3a46a3,fill opacity=0.5242,line cap=round,line join=round,line width=0.0185cm] (7.3321, 24.7408) -- (7.3814, 24.6593) -- (7.4297, 24.738);

  \path[draw=c3a46a3,fill opacity=0.5242,line cap=round,line join=round,line width=0.0185cm] (7.8513, 24.6228) -- (7.8513, 24.3735);

  \path[draw=c3a46a3,fill opacity=0.5242,line cap=round,line join=round,line width=0.0185cm] (7.8013, 24.4533) -- (7.8506, 24.3718) -- (7.8989, 24.4504);

  \path[draw=c3a46a3,fill opacity=0.5242,line cap=round,line join=round,line width=0.0185cm] (8.3833, 24.3114) -- (8.3833, 24.0621);

  \path[draw=c3a46a3,fill opacity=0.5242,line cap=round,line join=round,line width=0.0185cm] (8.3334, 24.1419) -- (8.3826, 24.0604) -- (8.431, 24.1391);

  \path[draw=c3a46a3,fill opacity=0.5242,line cap=round,line join=round,line width=0.0185cm] (8.915, 24.0058) -- (8.915, 23.7565);

  \path[draw=c3a46a3,fill opacity=0.5242,line cap=round,line join=round,line width=0.0185cm] (8.865, 23.8363) -- (8.9143, 23.7548) -- (8.9627, 23.8335);

  \path[draw=c3a46a3,fill opacity=0.5242,line cap=round,line join=round,line width=0.0185cm] (9.4426, 23.7093) -- (9.4426, 23.46);

  \path[draw=c3a46a3,fill opacity=0.5242,line cap=round,line join=round,line width=0.0185cm] (9.3926, 23.5398) -- (9.4419, 23.4583) -- (9.4902, 23.5369);

  \path[draw=c3a46a3,fill opacity=0.5242,line cap=round,line join=round,line width=0.0185cm] (9.9717, 23.4056) -- (9.9717, 23.1564);

  \path[draw=c3a46a3,fill opacity=0.5242,line cap=round,line join=round,line width=0.0185cm] (9.9217, 23.2362) -- (9.9709, 23.1547) -- (10.0193, 23.2333);

  \path[draw=c3a46a3,fill opacity=0.5242,line cap=round,line join=round,line width=0.0185cm] (10.4576, 23.1102) -- (10.4576, 22.8609);

  \path[draw=c3a46a3,fill opacity=0.5242,line cap=round,line join=round,line width=0.0185cm] (10.4076, 22.9407) -- (10.4569, 22.8592) -- (10.5052, 22.9379);

  \path[draw=c3a46a3,fill opacity=0.5242,line cap=round,line join=round,line width=0.0185cm] (10.9925, 22.8114) -- (10.9925, 22.5621);

  \path[draw=c3a46a3,fill opacity=0.5242,line cap=round,line join=round,line width=0.0185cm] (10.9425, 22.6419) -- (10.9918, 22.5604) -- (11.0401, 22.6391);

  \path[draw=c3a46a3,fill opacity=0.5242,line cap=round,line join=round,line width=0.0185cm] (8.6856, 27.3074) -- (8.6856, 27.0582);

  \path[draw=c3a46a3,fill opacity=0.5242,line cap=round,line join=round,line width=0.0185cm] (8.6356, 27.1379) -- (8.6849, 27.0564) -- (8.7332, 27.1351);

  \path[draw=c3a46a3,fill opacity=0.5242,line cap=round,line join=round,line width=0.0185cm] (9.1785, 27.0091) -- (9.1785, 26.7598);

  \path[draw=c3a46a3,fill opacity=0.5242,line cap=round,line join=round,line width=0.0185cm] (9.1285, 26.8396) -- (9.1778, 26.7581) -- (9.2261, 26.8367);

  \path[draw=c3a46a3,fill opacity=0.5242,line cap=round,line join=round,line width=0.0185cm] (9.6413, 26.7578) -- (9.6413, 26.5085);

  \path[draw=c3a46a3,fill opacity=0.5242,line cap=round,line join=round,line width=0.0185cm] (9.5913, 26.5883) -- (9.6406, 26.5068) -- (9.6889, 26.5855);

  \path[draw=c3a46a3,fill opacity=0.5242,line cap=round,line join=round,line width=0.0185cm] (10.1584, 26.4749) -- (10.1584, 26.2256);

  \path[draw=c3a46a3,fill opacity=0.5242,line cap=round,line join=round,line width=0.0185cm] (10.1084, 26.3054) -- (10.1577, 26.2239) -- (10.206, 26.3025);

  \path[draw=c3a46a3,fill opacity=0.5242,line cap=round,line join=round,line width=0.0185cm] (10.6276, 26.1873) -- (10.6276, 25.938);

  \path[draw=c3a46a3,fill opacity=0.5242,line cap=round,line join=round,line width=0.0185cm] (10.5776, 26.0178) -- (10.6269, 25.9363) -- (10.6752, 26.015);

  \path[draw=c3a46a3,fill opacity=0.5242,line cap=round,line join=round,line width=0.0185cm] (11.1597, 25.876) -- (11.1597, 25.6267);

  \path[draw=c3a46a3,fill opacity=0.5242,line cap=round,line join=round,line width=0.0185cm] (11.1097, 25.7065) -- (11.1589, 25.625) -- (11.2073, 25.7036);

  \path[draw=c3a46a3,fill opacity=0.5242,line cap=round,line join=round,line width=0.0185cm] (11.6913, 25.5704) -- (11.6913, 25.3211);

  \path[draw=c3a46a3,fill opacity=0.5242,line cap=round,line join=round,line width=0.0185cm] (11.6413, 25.4009) -- (11.6906, 25.3194) -- (11.739, 25.398);

  \path[draw=c3a46a3,fill opacity=0.5242,line cap=round,line join=round,line width=0.0185cm] (12.2189, 25.2738) -- (12.2189, 25.0246);

  \path[draw=c3a46a3,fill opacity=0.5242,line cap=round,line join=round,line width=0.0185cm] (12.1689, 25.1043) -- (12.2182, 25.0228) -- (12.2666, 25.1015);

  \path[draw=c3a46a3,fill opacity=0.5242,line cap=round,line join=round,line width=0.0185cm] (12.748, 24.9702) -- (12.748, 24.7209);

  \path[draw=c3a46a3,fill opacity=0.5242,line cap=round,line join=round,line width=0.0185cm] (12.698, 24.8007) -- (12.7473, 24.7192) -- (12.7956, 24.7979);

  \path[draw=c3a46a3,fill opacity=0.5242,line cap=round,line join=round,line width=0.0185cm] (13.2339, 24.6748) -- (13.2339, 24.4255);

  \path[draw=c3a46a3,fill opacity=0.5242,line cap=round,line join=round,line width=0.0185cm] (13.1839, 24.5053) -- (13.2332, 24.4238) -- (13.2815, 24.5024);

  \path[draw=c3a46a3,fill opacity=0.5242,line cap=round,line join=round,line width=0.0185cm] (13.7688, 24.376) -- (13.7688, 24.1267);

  \path[draw=c3a46a3,fill opacity=0.5242,line cap=round,line join=round,line width=0.0185cm] (13.7188, 24.2065) -- (13.7681, 24.125) -- (13.8164, 24.2036);

  \path[draw=black,line cap=butt,line join=miter,line width=0.0141cm] (8.3201, 20.1046) -- (9.6413, 20.8674);

  \path[draw=black,line cap=butt,line join=miter,line width=0.0141cm] (8.8863, 20.8674) -- (9.6413, 20.4315);

  \path[draw=black,line cap=butt,line join=miter,line width=0.0141cm] (8.3201, 20.3226) -- (8.3201, 20.1046) -- (8.6976, 20.1046);

  \node[text=black,line cap=butt,line join=miter,line width=0.0231cm,anchor=south west] (text9385-2-7-3) at (7.9126, 19.8262){N};

  \path[draw=black,line cap=butt,line join=miter,line width=0.0123cm,miter limit=4.0] (4.2727, 23.9952) -- (4.6502, 24.2132);

  \path[draw=black,line cap=butt,line join=miter,line width=0.0123cm,miter limit=4.0] (4.4614, 24.1042) -- (12.7659, 19.3096);

  \path[draw=c3a46a3,fill opacity=0.5032,line cap=round,line join=round,line width=0.0923cm] (11.7758, 22.4884) -- (12.3306, 22.8026) -- (12.208, 22.8776);

  \path[draw=black,line cap=butt,line join=miter,line width=0.0123cm,miter limit=4.0] (16.4356, 23.4877) -- (16.813, 23.7057);

  \path[draw=black,line cap=butt,line join=miter,line width=0.0123cm,miter limit=4.0] (12.5826, 19.207) -- (12.9601, 19.4249);

  \path[draw=black,line cap=butt,line join=miter,line width=0.0123cm,miter limit=4.0] (16.6243, 23.5967) -- (16.6243, 21.6353);

  \path[draw=black,line cap=butt,line join=miter,line width=0.0123cm,miter limit=4.0] (16.4356, 21.5263) -- (16.813, 21.7442);

  \path[draw=black,line cap=butt,line join=miter,line width=0.0123cm,miter limit=4.0] (16.4356, 21.3084) -- (16.813, 21.0904);

  \path[draw=black,line cap=butt,line join=miter,line width=0.0123cm,miter limit=4.0] (16.6243, 21.1994) -- (13.4158, 19.3469);

  \path[draw=black,line cap=butt,line join=miter,line width=0.0123cm,miter limit=4.0] (13.227, 19.4559) -- (13.6045, 19.238);

  \node[text=black,line cap=butt,line join=miter,line width=0.0231cm,anchor=south west] (text9385-2-7-36) at (15.1533, 19.8943){$8.0$m};

  \node[text=black,line cap=butt,line join=miter,line width=0.0231cm,anchor=south west] (text9385-2-7-36-5) at (16.7582, 22.7108){$3.2$m};

  \node[text=black,line cap=butt,line join=miter,line width=0.0231cm,anchor=south west] (text9385-2-7-36-5-5) at (7.3611, 21.5252){$20.1$m};

  \path[draw=cc2c2db,line cap=round,line join=round,line width=0.1845cm] (7.36, 26.5612) -- (7.36, 26.2886);

  \path[draw=cc2c2db,fill=cc2c2db,line cap=round,line join=round,line width=0.1028cm] (7.3625, 26.0587) ellipse (0.0497cm and 0.0113cm);

  \path[draw=cc2c2db,line cap=round,line join=round,line width=0.1845cm] (9.1595, 25.4917) -- (9.1595, 25.2192);

  \path[draw=cc2c2db,fill=cc2c2db,line cap=round,line join=round,line width=0.1028cm] (9.162, 24.9892) ellipse (0.0497cm and 0.0113cm);

  \path[draw=cc2c2db,line cap=round,line join=round,line width=0.1845cm] (10.67, 24.6037) -- (10.67, 24.3311);

  \path[draw=cc2c2db,fill=cc2c2db,line cap=round,line join=round,line width=0.1028cm] (10.6725, 24.1012) ellipse (0.0497cm and 0.0113cm);

  \path[draw=cc2c2db,line cap=round,line join=round,line width=0.1845cm] (6.0425, 27.9024) -- (6.0467, 27.3186) -- (12.2355, 23.7559);

  \path[draw=cc2c2db,line cap=round,line join=round,line width=0.1845cm] (12.2224, 23.7297) -- (12.2224, 23.4571);

  \path[draw=cc2c2db,fill=cc2c2db,line cap=round,line join=round,line width=0.1028cm] (12.2248, 23.2272) ellipse (0.0497cm and 0.0113cm);

  \path[draw=black,even odd rule,line cap=butt,line join=miter,line width=0.0154cm,miter limit=4.0] (12.1201, 22.7473) -- (12.1201, 22.9653) -- (12.6864, 23.2922);

  \path[draw=black,even odd rule,line cap=butt,line join=miter,line width=0.0154cm,miter limit=4.0] (12.1201, 22.9653) -- (11.3652, 23.4011);

\end{tikzpicture}

%% file: plenum.tex
\definecolor{cb10000}{RGB}{177,0,0}
\definecolor{cedecec}{RGB}{237,236,236}
\definecolor{ccec8c8}{RGB}{206,200,200}
\definecolor{cb4b3c5}{RGB}{180,179,197}
\definecolor{c302a4d}{RGB}{48,42,77}

\def \globalscale {1.000000}
\begin{tikzpicture}[y=1cm, x=1cm, yscale=\globalscale,xscale=\globalscale, every node/.append style={scale=\globalscale}, inner sep=0pt, outer sep=0pt]
  \path[draw=cb10000,fill=cedecec,fill opacity=0.5986,line cap=round,line join=round,line width=0.0209cm,dash pattern=on 0.1257cm off 0.0419cm] (6.8501, 24.0439) rectangle (14.656, 21.7439);

  \path[draw=cb10000,fill=cedecec,fill opacity=0.5986,line cap=round,line join=round,line width=0.0212cm,dash pattern=on 0.1273cm off 0.0424cm] (6.856, 26.5439) rectangle (14.656, 24.2439);

  \path[fill=white,line cap=round,line join=round,line width=0.0212cm,dash pattern=on 0.127cm off 0.0423cm] (8.256, 27.6439) rectangle (9.256, 24.1439);

  \path[fill=ccec8c8,fill opacity=0.6399,line cap=round,line join=round,line width=0.0212cm,dash pattern=on 0.127cm off 0.0423cm] (6.8507, 24.0232) rectangle (14.6612, 23.4439);

  \path[fill=ccec8c8,fill opacity=0.6399,line cap=round,line join=round,line width=0.0139cm,dash pattern=on 0.0835cm off 0.0278cm] (11.506, 25.1263) rectangle (14.006, 24.3439);

  \path[draw=black,line cap=round,line join=round,line width=0.0282cm] (8.256, 26.6439) -- (6.756, 26.6439) -- (6.756, 21.6439) -- (14.756, 21.6439) -- (14.756, 26.6616);

  \path[draw=black,line cap=round,line join=round,line width=0.0282cm] (14.756, 26.6616) -- (13.256, 26.6616);

  \path[draw=black,line cap=round,line join=round,line width=0.0282cm] (12.256, 26.6616) -- (9.256, 26.6616);

  \path[draw=black,line cap=round,line join=round,line width=0.0282cm] (8.256, 24.1439) -- (8.256, 27.1439);

  \path[draw=black,line cap=round,line join=round,line width=0.0282cm] (9.256, 27.1439) -- (9.256, 24.1439);

  \path[draw=black,line cap=round,line join=round,line width=0.0282cm] (12.256, 27.1439) -- (12.256, 26.6616);

  \path[draw=black,line cap=round,line join=round,line width=0.0282cm] (13.256, 27.1439) -- (13.256, 26.6616);

  \path[draw=black,line cap=round,line join=round,line width=0.0282cm] (12.256, 25.1439) -- (11.506, 25.1439) -- (11.506, 24.3439) -- (12.506, 24.3439) -- (12.506, 24.1439) -- (10.756, 24.1439);

  \path[draw=black,line cap=round,line join=round,line width=0.0282cm] (13.256, 25.1439) -- (14.006, 25.1439) -- (14.006, 24.3439) -- (13.006, 24.3439) -- (13.006, 24.1439) -- (14.756, 24.1439);

  \path[draw=black,line cap=round,line join=round,line width=0.0282cm] (12.256, 25.1439) -- (12.256, 25.5439);

  \path[draw=black,line cap=round,line join=round,line width=0.0282cm] (13.256, 25.1439) -- (13.256, 25.5439);

  \path[draw=black,line cap=round,line join=round,line width=0.0282cm] (10.256, 24.1439) -- (9.256, 24.1439);

  \path[draw=black,line cap=round,line join=round,line width=0.0282cm] (8.256, 24.1439) -- (6.756, 24.1439);

  \path[draw=black,line cap=round,line join=round,line width=0.0282cm] (10.256, 24.1439) -- (10.256, 24.4439);

  \path[draw=black,line cap=round,line join=round,line width=0.0282cm] (10.756, 24.1439) -- (10.756, 24.4439);

  \path[draw=black,fill=cb4b3c5,line cap=round,line join=round,line width=0.0212cm] (8.7509, 26.4618).. controls (8.8065, 26.4606) and (9.2061, 26.5712) .. (9.2426, 26.4618).. controls (9.2983, 26.2946) and (8.7957, 26.4362) .. (8.7509, 26.4618) -- cycle;

  \path[draw=black,fill=cb4b3c5,line cap=round,line join=round,line width=0.0212cm] (8.756, 26.4363).. controls (8.7004, 26.4376) and (8.3007, 26.327) .. (8.2643, 26.4363).. controls (8.2086, 26.6035) and (8.7111, 26.4619) .. (8.756, 26.4363) -- cycle;

  \path[draw=black,fill=cb4b3c5,line cap=round,line join=round,line width=0.0188cm] (8.7342, 26.4509) circle (0.0621cm);

  \path[draw=black,line cap=round,line join=round,line width=0.0106cm] (8.3042, 26.4714).. controls (8.3748, 26.496) and (8.5113, 26.4791) .. (8.5883, 26.4637);

  \path[draw=black,line cap=round,line join=round,line width=0.0106cm] (8.88, 26.4456).. controls (8.9891, 26.423) and (9.1137, 26.3847) .. (9.2177, 26.4159);

  \path[draw=black,line cap=round,line join=round,line width=0.0106cm,dash pattern=on 0.0635cm off 0.0212cm] (8.7034, 26.3026)arc(265.497:184.2084:0.4252 and 0.1417)arc(184.2084:102.9198:0.4252 and 0.1417)arc(102.9198:21.6312:0.4252 and 0.1417)arc(21.6312:-59.65730000000002:0.4252 and 0.1417);

  \path[draw=black,line cap=round,line join=round,line width=0.0106cm] (8.7034, 26.3026).. controls (8.7031, 26.3036) and (8.7027, 26.3045) .. (8.7034, 26.3026) -- cycle;

  \path[draw=black,line cap=round,line join=round,line width=0.0106cm] (8.6098, 26.273).. controls (8.6098, 26.2744) and (8.7034, 26.3026) .. (8.7034, 26.3026) -- (8.6264, 26.338);

  \path[draw=black,fill=cb4b3c5,line cap=round,line join=round,line width=0.0212cm] (12.7509, 25.2695).. controls (12.8065, 25.2682) and (13.2061, 25.3788) .. (13.2426, 25.2695).. controls (13.2983, 25.1022) and (12.7957, 25.2438) .. (12.7509, 25.2695) -- cycle;

  \path[draw=black,fill=cb4b3c5,line cap=round,line join=round,line width=0.0212cm] (12.756, 25.2439).. controls (12.7004, 25.2452) and (12.3007, 25.1346) .. (12.2643, 25.2439).. controls (12.2086, 25.4112) and (12.7111, 25.2696) .. (12.756, 25.2439) -- cycle;

  \path[draw=black,fill=cb4b3c5,line cap=round,line join=round,line width=0.0188cm] (12.7342, 25.2585) circle (0.0621cm);

  \path[draw=black,line cap=round,line join=round,line width=0.0106cm] (12.3042, 25.279).. controls (12.3748, 25.3037) and (12.5113, 25.2868) .. (12.5883, 25.2714);

  \path[draw=black,line cap=round,line join=round,line width=0.0106cm] (12.88, 25.2532).. controls (12.9891, 25.2307) and (13.1137, 25.1923) .. (13.2177, 25.2235);

  \path[draw=black,line cap=round,line join=round,line width=0.0106cm,dash pattern=on 0.0635cm off 0.0212cm] (12.7034, 25.1103)arc(265.497:184.2084:0.4252 and 0.1417)arc(184.2084:102.9199:0.4252 and 0.1417)arc(102.9198:21.6312:0.4252 and 0.1417)arc(21.6312:-59.657399999999996:0.4252 and 0.1417);

  \path[draw=black,line cap=round,line join=round,line width=0.0106cm] (12.7034, 25.1103).. controls (12.7031, 25.1112) and (12.7027, 25.1122) .. (12.7034, 25.1103) -- cycle;

  \path[draw=black,line cap=round,line join=round,line width=0.0106cm] (12.6098, 25.0807).. controls (12.6098, 25.082) and (12.7034, 25.1103) .. (12.7034, 25.1103) -- (12.6264, 25.1456);

  \path[draw=black,fill=white,fill opacity=0.5986,line cap=round,line join=round,line width=0.0141cm] (12.856, 27.2439) -- (12.856, 26.2439) -- (12.956, 26.2439) -- (12.756, 26.0439) -- (12.556, 26.2439) -- (12.656, 26.2439) -- (12.6641, 26.7439).. controls (12.6094, 26.3629) and (12.3965, 26.3035) .. (12.292, 26.2243) -- (12.3218, 26.1547) -- (12.1398, 26.2169) -- (12.2064, 26.3968) -- (12.2425, 26.3223).. controls (12.3363, 26.3416) and (12.556, 26.5401) .. (12.556, 26.7439) -- (12.556, 27.2439) -- cycle;

  \path[draw=black,fill=white,fill opacity=0.5986,line cap=round,line join=round,line width=0.0141cm] (12.9126, 24.4596) -- (12.9126, 23.4596) -- (13.0126, 23.4596) -- (12.8126, 23.2596) -- (12.6126, 23.4596) -- (12.7126, 23.4596) -- (12.7207, 23.9596).. controls (12.7107, 23.6667) and (12.483, 23.6337) .. (12.3041, 23.6139) -- (12.3041, 23.5422) -- (12.1397, 23.6747) -- (12.2968, 23.7937) -- (12.2992, 23.7206).. controls (12.4877, 23.7336) and (12.6126, 23.8072) .. (12.6126, 23.9596) -- (12.6126, 24.4596) -- cycle;

  \path[draw=black,fill=white,fill opacity=0.6399,line cap=round,line join=round,line width=0.0141cm] (12.856, 25.9439) -- (12.856, 25.5439) -- (12.956, 25.5439) -- (12.756, 25.3439) -- (12.556, 25.5439) -- (12.656, 25.5439) -- (12.656, 25.6439).. controls (12.6474, 25.7672) and (12.5452, 25.7439) .. (12.0452, 25.7439) -- (12.0452, 25.8975).. controls (12.5452, 25.8975) and (12.656, 25.8975) .. (12.656, 25.7975) -- (12.656, 25.9439) -- cycle;

  \path[draw=black,fill opacity=0.6399,line cap=round,line join=round,line width=0.0141cm] (8.856, 26.7439) -- (8.856, 27.1439) -- (8.956, 27.1439) -- (8.756, 27.3439).. controls (8.756, 27.3439) and (8.556, 27.1439) .. (8.556, 27.1439).. controls (8.556, 27.1439) and (8.656, 27.1439) .. (8.656, 27.1439) -- (8.656, 26.7439) -- cycle;

  \path[draw=black,fill=white,fill opacity=0.7257,line cap=round,line join=round,line width=0.0141cm] (10.589, 23.9507) -- (10.589, 24.5148) -- (10.6585, 24.5148) -- (10.5196, 24.7154) -- (10.3806, 24.5148) -- (10.4501, 24.5148) -- (10.4501, 23.9507) -- cycle;

  \path[draw=black,fill=white,fill opacity=0.7257,line cap=round,line join=round,line width=0.0141cm] (9.6893, 23.2852) -- (9.6893, 23.5409) -- (9.8459, 23.5409) -- (9.4113, 23.72) -- (8.9575, 23.5409) -- (9.1332, 23.5409) -- (9.1332, 23.2852) -- cycle;

  \path[draw=c302a4d,fill opacity=0.7257,draw opacity=0.6037,line cap=round,line join=round,line width=0.0141cm] (10.096, 25.8177)arc(143.3895:75.7071:0.3301)arc(75.7072:8.0249:0.3301)arc(8.0248:-59.65750000000003:0.3301);

  \path[draw=c302a4d,fill opacity=0.7257,draw opacity=0.6037,line cap=round,line join=round,line width=0.0141cm] (10.2455, 25.4349)arc(295.521:227.6359:0.4653)arc(227.6359:159.7509:0.4653)arc(159.7508:91.8657:0.4653)arc(91.8657:23.9807:0.4653);

  \path[draw=c302a4d,fill opacity=0.7257,draw opacity=0.6037,line cap=round,line join=round,line width=0.0141cm] (9.3618, 25.321)arc(146.4218:78.4438:0.1948)arc(78.4437:10.4657:0.1948)arc(10.4657:-57.51220000000001:0.1948)arc(302.4877:234.5097:0.1948);

  \path[draw=c302a4d,fill opacity=0.7257,draw opacity=0.6037,line cap=round,line join=round,line width=0.0141cm] (9.4722, 25.3175) -- (9.3618, 25.321) -- (9.4033, 25.4372);

  \path[draw=c302a4d,fill opacity=0.7257,draw opacity=0.6037,line cap=round,line join=round,line width=0.0141cm] (10.1819, 25.8213) -- (10.096, 25.8177) -- (10.1581, 25.8812) -- (10.096, 25.8177) -- (10.089, 25.9169);

  \path[draw=c302a4d,fill opacity=0.7257,draw opacity=0.6037,line cap=round,line join=round,line width=0.0141cm] (10.3901, 26.0929) -- (10.5231, 26.0439) -- (10.5301, 26.1795);

  \path[draw=c302a4d,fill opacity=0.7257,draw opacity=0.6037,line cap=round,line join=round,line width=0.0141cm] (11.2905, 24.9394)arc(319.1623:230.5532:0.2374)arc(230.5531:141.944:0.2374)arc(141.9441:53.3349:0.2374);

  \path[draw=c302a4d,fill opacity=0.7257,draw opacity=0.6037,line cap=round,line join=round,line width=0.0141cm] (11.1736, 24.943) -- (11.2905, 24.9394) -- (11.2408, 24.8186);

  \path[draw=c302a4d,fill opacity=0.7257,draw opacity=0.6037,line cap=round,line join=round,line width=0.0141cm,cm={ -0.6882,-0.7256,-0.7256,0.6882,(21.5503, 9.2605)}] (-6.5108, 16.2497)arc(319.1623:230.5532:0.2374)arc(230.5531:141.9443:0.2374)arc(141.9442:53.335:0.2374);

  \path[draw=c302a4d,fill opacity=0.7257,draw opacity=0.6037,line cap=round,line join=round,line width=0.0141cm] (14.3173, 25.2555) -- (14.2395, 25.1681) -- (14.3613, 25.1211);

  \path[draw=c302a4d,fill opacity=0.7257,draw opacity=0.6037,line cap=round,line join=round,line width=0.0141cm] (9.7681, 24.4523)arc(265.4969:188.9628:0.2705)arc(188.9628:112.4287:0.2705)arc(112.4287:35.8946:0.2705)arc(35.8946:-40.6395:0.2705);

  \path[draw=c302a4d,fill opacity=0.7257,draw opacity=0.6037,line cap=round,line join=round,line width=0.0141cm] (9.967, 24.6517).. controls (9.9741, 24.6517) and (9.9947, 24.5458) .. (9.9947, 24.5458) -- (10.1248, 24.5851);

  \node[text=black,anchor=south west,line cap=round,line join=round,line width=0.0141cm] (text44) at (6.9687, 21.8874){$C_0$};

  \node[text=black,anchor=south west,line cap=round,line join=round,line width=0.0141cm] (text44-2) at (6.9687, 23.5829){$C_0'$};

  \node[text=black,anchor=south west,line cap=round,line join=round,line width=0.0141cm] (text44-2-9) at (6.9687, 24.3742){$C_1$};

  \node[text=black,anchor=south west,line cap=round,line join=round,line width=0.0141cm] (text44-2-9-1) at (14.1141, 24.3742){$V_1$};

  \node[text=black,anchor=south west,line cap=round,line join=round,line width=0.0141cm] (text44-2-9-1-9) at (14.1141, 21.8874){$V_0$};

  \node[text=black,anchor=south west,line cap=round,line join=round,line width=0.0141cm] (text44-2-9-1-9-6) at (14.1141, 23.5163){$V_0'$};

  \node[text=black,anchor=south west,line cap=round,line join=round,line width=0.0141cm] (text44-2-9-1-9-6-6) at (13.5442, 24.4564){$V_1'$};

  \node[text=black,anchor=south west,line cap=round,line join=round,line width=0.0141cm] (text44-2-9-1-9-1) at (8.6397, 27.4737){$q$};

  \node[text=black,anchor=south west,line cap=round,line join=round,line width=0.0141cm] (text44-2-9-1-9-1-4) at (12.6397, 27.4331){$q$};

  \node[text=black,anchor=south west,line cap=round,line join=round,line width=0.0141cm] (text44-2-9-1-9-1-4-3) at (10.7867, 25.699){$\gamma_1 q+q'$};

  \node[text=black,anchor=south west,line cap=round,line join=round,line width=0.0141cm] (text44-2-9-1-9-1-4-3-1) at (12.9792, 25.9108){$(1-\gamma_1)q$};

  \node[text=black,anchor=south west,line cap=round,line join=round,line width=0.0141cm] (text44-2-9-1-9-1-4-3-1-6) at (11.4495, 26.1732){$\gamma_1q$};

  \node[text=black,anchor=south west,line cap=round,line join=round,line width=0.0141cm] (text44-2-9-1-9-1-4-8) at (10.4409, 24.8156){$q'$};

  \node[text=black,anchor=south west,line cap=round,line join=round,line width=0.0141cm] (text44-2-9-1-9-1-4-8-5) at (12.6079, 24.5601){$Q$};

  \node[text=black,anchor=south west,line cap=round,line join=round,line width=0.0141cm] (text44-2-9-1-9-1-4-8-5-0) at (12.5952, 22.9773){$\gamma_0 Q$};

  \node[text=black,anchor=south west,line cap=round,line join=round,line width=0.0141cm] (text44-2-9-1-9-1-4-8-5-0-4) at (9.1796, 22.9025){$\gamma_0 Q$};

  \node[text=black,anchor=south west,line cap=round,line join=round,line width=0.0141cm] (text44-2-9-1-9-1-4-8-5-0-1) at (10.6615, 23.5653){$(1-\gamma_0) Q$};

  \node[text=black,anchor=south west,line cap=round,line join=round,line width=0.0141cm] (text44-2-9-3) at (11.5815, 24.4337){$C_1'$};

\end{tikzpicture}